\documentclass[12pt,preprint]{aastex}

\usepackage{lscape,graphicx}
\usepackage{epstopdf}

\def\deg      {{\ifmmode^\circ\else$^\circ$\fi}}

  \shorttitle{Is there a need for dry mergers?}
 \shortauthors{Scarlata et al.}
  \begin{document}
  \title{The redshift evolution of early-type galaxies in COSMOS: \\
    Do massive early-type galaxies form by dry mergers?}  \author{ C.
    Scarlata\altaffilmark{1}, C. M.  Carollo,\altaffilmark{1} S.J.
    Lilly\altaffilmark{1}, R.  Feldmann\altaffilmark{1}, P.
    Kampczyk\altaffilmark{1}, A.  Renzini\altaffilmark{2}, A.
    Cimatti\altaffilmark{3}, C.  Halliday\altaffilmark{3}, E.
    Daddi\altaffilmark{4}, M. T.  Sargent\altaffilmark{1}, A.
    Koekemoer\altaffilmark{5}, N.  Scoville\altaffilmark{6}, J-P.
    Kneib\altaffilmark{7}, A.  Leauthaud\altaffilmark{7}, R.
    Massey\altaffilmark{6}, J.  Rhodes\altaffilmark{6,8}, L.
    Tasca\altaffilmark{7}, Capak\altaffilmark{6}, H. J.
    McCracken\altaffilmark{9}, B.  Mobasher\altaffilmark{5}, Y.
    Taniguchi\altaffilmark{10}, D.  Thompson\altaffilmark{6,11}, M.
    Ajiki\altaffilmark{12}, H.  Aussel\altaffilmark{13,14}, T.
    Murayama\altaffilmark{12}, D. B.  Sanders\altaffilmark{13}, S.
    Sasaki\altaffilmark{12,15}, Y.  Shioya\altaffilmark{15}, M.
    Takahashi\altaffilmark{12,15}}

\altaffiltext{$\star$}{Based on observations with the NASA/ESA {\em
Hubble Space Telescope}, obtained at the Space Telescope Science
Institute, which is operated by AURA Inc, under NASA contract NAS
5-26555.}  

\altaffiltext{1}{Department of Physics, Swiss Federal Institute of
  Technology (ETH-Zurich), CH-8093 Zurich, Switzerland}

\altaffiltext{2}{Dipartimento di Astronomia, Universit\'a di Padova,
  vicolo dell'Osservatorio 2, I-35122 Padova, Italy}
  
\altaffiltext{3}{INAF- Osservatorio Astrofisico di Arcetri, Largo E.
  Fermi 5, I-50125 Firenze, Italy}
  
\altaffiltext{4}{National Optical Astronomy Observatory, P.O. Box
  26732, Tucson, AZ 85726}

\altaffiltext{5}{Space Telescope Science Institute, 3700 SanMartin
  Drive, Baltimore, MD 21218} 
  
\altaffiltext{6}{California Institute of Technology, MC 105-24, 1200
  East California Boulevard, Pasadena, CA 91125}

\altaffiltext{7}{Laboratoire d'Astrophysique de Marseille, BP 8,
  Traverse du Siphon, 13376 Marseille Cedex 12, France}

\altaffiltext{8}{Jet Propulsion Laboratory, Pasadena, CA 91109}

\altaffiltext{9}{Institut d'Astrophysique de Paris, UMR 7095, 98 bis
  Boulevard Arago, 75014 Paris, France}

\altaffiltext{10}{Subaru Telescope, National Astronomical Observatory
  of Japan, 650 North Aohoku Place, Hilo, HI 96720}

\altaffiltext{11}{Large Binocular Telescope Observatory,University of
  Arizona, 933 N. Cherry Ave., Tucson, AZ 85721}

\altaffiltext{12}{Astronomical Institute, Graduate School of Science,
  Tohoku University, Aramaki, Aoba, Sendai 980-8578, Japan}

\altaffiltext{13}{Institute for Astronomy, 2680 Woodlawn Dr.,
  University of Hawaii, Honolulu, Hawaii, 96822}

\altaffiltext{14}{Service d'Astrophysique, CEA/Saclay, 91191
  Gif-sur-Yvette, France}

\altaffiltext{15}{Physics Department, Graduate School of Science and
  Engineering, Ehime University, Japan}
 
\begin{abstract}
  We study the evolution since $z \sim 1$ of the rest-frame $B$
  luminosity function of the early-type galaxies (ETGs) in $\sim 0.7$
  degrees$^2$ in the COSMOS field. In order to identify {\it all}
  plausible progenitors of local ETGs we construct the sample of
  high$-z$ galaxies using two complementary criteria: $(i)$ A {\it
    morphological} selection based on the Zurich Estimator of
  Structural Types, and $(ii)$ A {\it photometric} selection based on
  the galaxy properties in the rest-frame $(U-V)$-$M_V$
  color-magnitude diagram. We furthermore constrain both samples so as
  to ensure that the selected high$-z$ progenitors of ETGs are
  compatible with evolving into systems which obey a fundamental $z=0$
  scaling relation for early-type galaxies, i.e., the $\mu_B-r_{hl}$
  Kormendy relation.  Assuming the luminosity evolution derived from
  studies of the fundamental plane for high$-z$ ETGs, our analysis
  shows no evidence for a decrease in the number density of the most
  massive ETGs out to $z\sim 0.7$: Both the morphologically- and the
  photometrically-selected sub-samples show no evolution in the number
  density of bright ($\approx L>2.5L^*$) ETGs. Allowing for different
  star formation histories, and cosmic variance, we estimate a maximum
  decrease in the number density of massive galaxies at that redshift
  of $\sim 30$\%.  \\ We observe, however, both in the photometrical
  and in the morphological samples, a deficit of up to $\sim 2-3$ of
  fainter early-type galaxies over the same cosmic period.  Our
  results argue against a significant contribution of recent
  dissipationless ``dry'' mergers to the formation of the most massive
  early-type galaxies. We suggest that the mass growth in low
  luminosity ETGs can be explained with a conversion from $z\sim0.7$
  to $z=0$ of blue, irregular and disk galaxies into low- and
  intermediate-mass ``red'', early-type galaxies, possibly also
  through gas rich mergers. This interpretation is consistent with the
  observed increase of a factor of the order of $\sim 2-3$, from $z=0$
  to $z=0.7$, of the rest-frame $B$-band luminosity function of  blue irregular galaxies.\\
 \end{abstract}
 
 
\section{Introduction}

One of the currently most debated -- and consequential -- issues in
astrophysics is the formation of massive elliptical galaxies
\citep[see][for a recent review]{renzini2006}.

From the theoretical point of view, CDM simulations can accurately
predict the redshift evolution of dark matter haloes. Specifically, on
large scales CDM simulations have enjoyed great success in accounting
for the growth through cosmic times of the structures - clusters,
filaments and voids - starting from the extremely smooth initial
conditions inferred from the cosmic microwave background
\citep[e.g.][]{springel2006}.  We still lack, however, an
understanding of the astrophysical processes which produce massive
galaxies with the properties that these show in the local Universe:
gas cooling, star formation, stellar and AGN feedback, are all poorly
-if at all- understood processes, which are currently parametrized
with {\it ad hoc} recipes in galaxy formation models.  Simulations
have been less successful on these lower, galactic scales ($\sim 100$
kpc) and it is not clear yet whether this is due to the just mentioned
complexities of baryon physics, or to other reasons.

From the observational point of view, the emerging picture is still
uncertain. In the local Universe, studies of the stellar populations
of massive ellipticals indicate a formation epoch for the bulk of
their stars at redshifts $z>2$; however, the old stellar ages at $z=0$
cannot break the degeneracy between the mass assembly of old smaller
sub-units relative to in-situ star formation
\citep{carollo1993,bender1993, carollo1994a,
  carollo1994b,bernardi2003d,thomas2005}.  This degeneracy can in
principle be removed by observations of the evolution of the number
density of ellipticals as a function of redshift; however, small
statistics, cosmic variance and details in the different analysis of
the high$-z$ samples have generated a debate as to whether massive
ellipticals are already fully assembled by $z\sim 1$ -- a fact which,
if true, would require considerable rethinking of the currently
favored galaxy formation models (see, e.g., Bell et al. 2004a, and
Daddi et al. 2005 for examples of different views, and Renzini 2006
for an extensive review and further references).

Some studies have suggested that the majority of the most massive
($M>M^*$) early-type galaxies (ETGs) is assembled over the last few
Gyr from the dissipationless (``dry'') mergers of less massive ETGs
\citep{bell2004a,faber2005}.  Estimates of the merger rate since
$z\sim 1$ have been attempted by several authors (e.g.  van Dokkum
2005; Bell et al 2006a; Lin et al. 2004; Kampczyk et al. 2006).
\citet{vandokkum2005} used the statistics of relict tidal interactions
in the local universe and conclude that $\sim 35\%$ of bulge-dominated
galaxies should have experienced a merger with mass ratio $>$ 1:4
since $z\sim 0.1$.  Based on 6 ETG-ETG close pairs in the GEMS field,
\citet{bell2006a} estimate that each present-day ETG with $M_{\rm
  v}<-20.5$ has undergone $\sim$0.5--2 major dry mergers since $z\sim
0.7$.  Several other studies however dispute such high frequencies of
dry mergers since $z\sim1$. For example, \citet{bell2006b}, based on
the 3D two-point correlation function of ETGs, conclude that only 20\%
of all $M>2.5\times 10^{10}$M$_{\odot}$ galaxies have experienced a
major merger since $z=0.8$. This result roughly agrees with the
analysis of Lin and collaborators who, from the count of close pairs
in the DEEP2 sample, conclude that only $\sim 9\%$ of present-day
$M^*$ galaxies have undergone a major merger over the same period.
Supporting evidence against a large contribution of dry mergers to the
formation of massive elliptical galaxies comes also from estimates of
the dry merging rate at $z< 0.36$ from the SDSS database
\citet{masjedi2006}, which show a $<1\%$ probability per Gyr for an
ETG to merge with another ETG.  The Masjedi et al.  estimate implies a
dry-merging rate much lower than the rate at which ETG-hosting dark
matter halos merge with one another \citep{hogg2006}, possibly
highlighting a potential problem with CDM simulations, or in the way
these simulations are related to the observed galaxy populations.

In any event, there are clearly major discrepancies among current
attempts at measuring the dry merger rate, which in part are the
result of the uncertainty affecting the time it will take to a given
galaxy pair to merge, or to some tidal debris to disappear, and in
part to small-number statistics. In this paper we follow a different
approach in order to set limits to the role played by dry mergers in
establishing the population of massive ETGs in the local universe.
Here we map the evolution with redshift up to $z\sim 0.7$ of the
number density of massive ETGs.  Since the bulk of stars in these
galaxies formed at much higher redshifts (see the extensive literature
quoted in Renzini 2006), an increase in the number density of massive
ETGs since $z=0.7$ should be ascribed to dry mergers, alternatively, a
lack of evidence for such an increase would allow us to set limits on
the dry merging rate.

We use the data from the COSMOS program \citep{scoville2006a} to study
the evolution of the luminosity function (LF) of the ETGs up to
redshift $z=1$. We base the analysis on the sample of $\sim 32,000$,
$I_{AB}\le 24$ COSMOS galaxies with reliable photometric redshift in
the range $0.2<z\le 1.0$ presented in \citet[][hereafter
Paper~I]{scarlata2006a} to extract a combined sample of 3980
morphologically- and/or photometrically-selected progenitors of
early-type systems to study, in the $z=0$ to $z=1$ redshift window,
their LF evolution in the rest-frame $B-$band.  Specifically, we
select two complementary samples of possible progenitors of massive
ETGs, and study their individual and combined redshift evolution.  The
first sample is selected morphologically, using the classification of
ZEST (Zurich Estimator of Structural Types, Paper~I).  The second
sample is selected photometrically, using the red-sequence identified,
at all redshifts, on the rest--frame $(U-V)-M_V$ diagram.  Both
samples are further constrained by requiring that, by
passive-evolution fading, the high$-z$ progenitors of ETGs evolve into
$z=0$ systems that lie on the $\mu_B-r_{hl}$ ``Kormendy relation''
\citep{kormendy1977}.

Our double (i.e., morphological and photometrical) selection is
motivated by the fact that, at $z=0$, elliptical galaxies are
characterized by well defined stellar population \underline{and}
dynamical/structural properties.  Specifically, $z=0$ ellipticals {\it
  both} $(a)$ Have uniformly--old stellar populations with a small
scatter in the observed colors, implying a small scatter in the
formation epoch of their stellar populations; {\it and} $(b)$ Are
described by regular surface brightness distributions, well
represented by almost deVaucouleurs density profiles, indicating a
large role of violent relaxation in their dynamical history.
Depending on the relative timing and importance of in-situ star
formation versus stellar mergers, $z\sim 1$ progenitors of ellipticals
may thus appear as already morphologically-relaxed systems or possibly
show passively-evolving stellar populations with irregular
morphologies due to recent mergers.  Therefore, to identify {\it all}
plausible high$-z$ progenitors of local ETGs we consider the sample
 created by the union of the morphologically and photometrically
selected galaxies. This sample allows us to be comprehensive in the
``counting'' of the number density of massive galaxies at earlier
epochs. Still, selection biases remain in our analysis: The
photometric-selected sample could include galaxies which are actually
not progenitors of ellipticals and have red colors simply because of
the effects dust extinction.  On the other hand, blue progenitors with
merger morphologies would not enter in either of the samples.  We
discuss later in the paper the impact of these biases on our
conclusions.

The paper is organized as follows. Section 2 describes the data and
the basic measurements; Section 3 describes in detail the sample
selection criteria; Section 4 quickly summarizes the structural
properties of the photometrically-selected ETGs, and the color
properties of the morphologically-selected sample; Section 5 presents
our main analysis, i.e., the evolution with redshift of the rest-frame
$B$-band LF of the different and also combined samples of high-$z$ ETG
progenitors. We discuss our results in Section 6 and highlight a few
concluding remarks in Section 7.  Appendix~A presents the tests
performed on the LFs to account for the photometric redshift
uncertainties, and Appendix~B presents the SDSS-based $z=0$
comparison sample.  Throughout the paper we assume $\Omega_m=0.25$,
$\Omega_m + \Omega_{\Lambda}=1$, and $H_{0}=70$ km s$^{-1}$
Mpc$^{-1}$. All magnitudes are AB magnitudes \citep{oke1974}, unless
otherwise specified.
 
\section{Data and basic measurements}
 
\subsection{The data and the input catalogue}

The analysis presented in this paper uses the COSMOS Hubble Space
Telescope (HST) Advanced Camera for Survey (ACS) F814W images
(hereafter $I_{AB}$), and ancillary ground-base UV to near-infrared
data available for the COSMOS field. The ground-based $B_J$, $g^{+}$,
$V_J$, $r^{+}$, $i^{+}$, $z^{+}$ data were acquired with the Subaru
Telescope, the $u^{*}$-band data with the Canada--France--Hawaii
Telescope, and the infrared $K_S$ images with the Cerra Tololo
International Observatory and Kitt Peak National Observatory
telescopes.  The observations and the data processing are described in
detail in \cite{capak2006}, \cite{tanaguchi2006}, for the ground-based
data; and in \cite{kokomero2006} for the HST-ACS data.

Photometric catalogs were created separately for the ground-based and
the HST data.  The ACS--based catalog was generated by
\cite{leauthaud2006} using SExtractor \citep{sextractor} in a two
steps strategy, in order to correctly detect and deblend objects with
a wide range of magnitudes and sizes.
Extensive simulations presented in \cite{leauthaud2006}
show that the ACS catalog is at least 90\% complete for objects
smaller than 1\farcs4 and $I_{AB}\le 24$  (roughly corresponding 
to a surface brightness limit of $\mu_{I_{AB}}=25$ mag arcsec$^{-2}$).

The ground-based catalog was generated using SExtractor in dual-image
mode. The detection was performed in the original best--seeing PSF
$i^+-$band image, while the photometry was measured in the PSF-matched
images.  This process ensured that the photometry of nearby galaxies
was optimally deblended in the final catalog.  Magnitudes were
measured within apertures of $3''$ diameter; and $5\sigma$ magnitude
limits are 26.4, 27.3, 27.0, 26.6, 26.8, 26.2, 25,2, and 21.6 in the
$u^{*}$, $B_J$, $g^{+}$, $V_J$, $r^{+}$, $i^{+}$, $z^{+}$, and $K_S$
filters, respectively \citep{capak2006}.

In this work, we present the results for the central area of the
COSMOS field of $\sim$0.74 degrees$^2$ covered by the 260 ACS
pointings acquired during HST Cycle~12 observations. We limit our
study to the $\sim 32,000$ galaxies with total magnitude (as measured
by the SExtractor MAG\_AUTO) brighter than $I_{AB}=24.0$ and
photometric redshifts (see next Section) in the redshift range
0.2$<z\le 1.0$.

\subsection{Photometric redshifts}
\label{sec:photoz}

We adopt the Maximum Likelihood photometric redshift estimates
obtained for the COSMOS galaxies by \citet{feldmann2006} with the
Zurich Extragalactic Bayesian Redshift Analyzer (ZEBRA). Our photo$-z$
were obtained using as a basic set of galaxy templates the empirical
spectra by \citet{coleman1980} and \citet{kinney1996}, that cover the
spectral types from elliptical to the star-forming galaxies. By means
of an iterative technique ZEBRA automatically corrects their original
set of templates to best represent the galaxy spectral energy
distributions (SEDs) in different redshift bins (to empirically take
into account, e.g., dust absorption effects and other possible
inadequacies inherent in the original set of templates).  The
availability of a ``training set'' of spectroscopically--derived
$z-$COSMOS redshifts \citep{lilly2006} for a small fraction of the
whole photometric sample under investigation allows ZEBRA to achieve
an optimal correction of the galaxy templates and a precise
calibration of the photometric redshifts, and thus accurate photo$-z$
estimates. 

The resulting photometric redshifts have an accuracy of $\sigma_z/
(1+z)= 0.027$ over the whole redshift range considered in the current
analysis, relative to the $z$COSMOS spectroscopic redshifts of
$I_{AB}\le 22.5$ galaxies.  Tests on the $I_{AB} \le 22.5$
spectroscopic sample artificially fainted to reproduce a $I_{AB} \le
24$ sample, indicate photometric redshift errors $\sigma_z/ (1+z)\sim
0.06$.

We checked that our main results remain unchanged when: $(a)$ changing
the redshift binning adopted by ZEBRA for the template optimization,
the minimum errors in the photometry allowed by ZEBRA, the corrections
applied to the photometric catalog (i.e., the pliantness parameter
$-\sigma-$ and the regularization parameter $-\rho-$, that appear in
the $\chi^2$ minimization approach of the ZEBRA code; Feldman et al.
2006); $(b)$ using only galaxies with $\chi^2< \chi^2_{95}$ in the
ZEBRA fits, with $\chi^2_{95}$ the value corresponding to the
95$^{th}$ percentile in an ideal $\chi^2$-distribution of a varying
number of degrees of freedom; or all galaxies in the sample,
independent of their $\chi^2$ in the ZEBRA photometric fits.

For a total of $\sim 4.4\%$ of the galaxies that were detected in the
original ACS--based catalogue it was not possible to derive
photometric redshift estimate, due to either their absence, or their
blending with other galaxies in the ground-based catalogue. These
galaxies were excluded from our analysis. The usable Cycle~12
$I_{AB}\le 24$ ACS-based catalogue contains 32540 galaxies with
photometric redshift in the range [0.2,1.0] and $\chi^2< \chi^2_{95}$.

\subsection{The ZEST morphological classification }
\label{sec:morphologyselection}

To extract the sample of {\it morphologically-selected} ETGs at high
redshifts, and also to study later on in the paper the morphologies of
the photometrically--selected sample of ETGs, we used the ZEST
classification presented in Paper~I.

ZEST quantitatively describes the galaxy structure using three
variables ($PC_{1}$; $PC_2$; $PC_3$), obtained by performing a
principal component analysis (PCA) in the five-dimensional parameter
space of asymmetry ($A$), concentration ($C$), Gini coefficient ($G$),
the 2nd--order moment of the brightest 20\% galaxy pixels
\citep[M$_{20}$, e.g.,][]{abraham2003,lotz2004}, and the ellipticity
of the light distribution\footnote{In order to ensure a meaningful
  comparison of the parameters among galaxies at different redshifts,
  all coefficients are computed within elliptical apertures of
  semi-major axis equal to one Petrosian radius
  \citep[][]{petrosian1976}} ($\epsilon$). These non--parametric
diagnostics provide complementary, but also redundant information on
galaxy structure.  With the PCA we found that the first three
$PC$ variables explain more than 90\% of the variance in the
original dataset, and thus completely describe the galaxy structure.

The morphological classification is performed in the space with axes
$PC_{1}-PC_2-PC_3$. To associate a (dominant) morphological class to
different regions in the $PC-$space, the latter was partitioned into a
regular 3D-grid with unit steps in each of the coordinates, and all
the galaxies in the COSMOS sample within each of the unit $PC$-cubes
were visually inspected. Each galaxy was then assigned the
morphological class of the $PC-$cube corresponding to its position.
The ZEST classification associates to each $PC-$cube a Type ($=1$ for
early-type galaxies; $=2$ for disk galaxies; and $=3$ for irregular
galaxies).  Furthermore, each $PC-$cube classified as Type~2 is
assigned a ``bulgeness'' parameter according to the median value
($n_{\rm m}$) of the distribution of $n$ Sersic indices of all
galaxies brighter than $I_{AB}=22.5$ in that cube. The two-dimensional
GIM2D fits were performed by \citet{sargent2006} on the COSMOS-bright
sample, i.e., $I_{AB}\le 22.5$; we verified in Paper~I that the PCA
provides consistent results if separately applied to the faint (i.e.,
$I_{AB}\le 24$) and the bright (i.e., $I_{AB}\le 22.5$) samples. The
``bulgeness'' parameter is related to the $B/D$ ratio \footnote{When
  single Sersic fits are performed to bulge$+$disk galaxies, the $n$
  Sersic index of the {\it global} galaxy profile is found to be
  monotonically related to the galaxy $B/D$ ratio
  \citep[e.g.,][]{blanton2003}.} (see Paper~I for detail) and ranges
from 0 (bulge dominated disk galaxies, Type~2.0) to 3 (pure disk
galaxies, Type~2.3). We use $n_{\rm m}$ to assign the bulgeness
parameter to each Type~2-classified PC-cube, according to the following
scheme: bulgeness $=3,2,1,$ and $0$ respectively for $0<n_{\rm m}<
0.75$, $0.75\le n_{\rm m} < 1.25$, $1.25 \le n_{\rm m} < 2.5$, and
$n_{\rm m} \ge 2.5$. The distributions of Sersic $n$ for all galaxies
classified as Type~2, splitted according to their bulgeness parameter
are shown in Figure~9 of Sargent et al. (2007).

In Paper~I we discuss in detail the uncertainties and the systematic
errors in the measured structural parameters as a function of
signal-to-noise ratio (S/N), and to what extent the COSMOS-calibrated
ZEST morphological classification grid is affected by the S/N of the
individual galaxies; our tests show that the ZEST morphological
classification is robust down to $I_{AB}=24$.  Furthermore, the ZEST
classification is substantially more efficient in disentangling
different galactic types than simpler classification schemes based on,
e.g., a threshold in $n$-Sersic index, or two/three of the
non-parametric diagnostics mentioned above.

\subsection{Rest-frame magnitudes and colors}

In order to compute the rest--frame absolute $B-$band magnitude
($M_B$), we define a color $k-$correction $K_{BI}(z)=B -I_{z}$, where
$B$ is the magnitude at $\lambda _{B,{\rm obs}}=(1+z)\times\lambda
_B$, and $I_{z}$ indicates the observed $I-$band magnitude for a
galaxy at redshift $z$. The central wavelength of the ACS filter
(F814W) corresponds to the central wavelength of the $B-$band at
redshift $z\sim 0.8$, therefore $B-I_{z\sim0.8}=0$.  It follows that
the rest--frame absolute magnitude $M_B$ can be expressed as:

\begin{equation}
  M_B=I_{z} -5.0\log{(d_L(z)/10{\rm pc})}-25.0 + K_{BI}(z) +2.5\log{(1+z)},
\end{equation}

\noindent
where $d_L$ is the luminosity distance at redshift $z$.  The color $B
-I_{z}$ depends on the galaxy spectral energy distribution. For each
galaxy we therefore use its ZEBRA best--fit template and the
corresponding photometric redshift to compute the $B -I_{z}$ color.

The rest--frame $U-V$ colors are derived from the ground-based
photometry, by interpolating the observed SEDs at the wavelengths of
the $U$ and $V$ filters, redshifted at the galaxy photometric
redshift. The adopted effective wavelengths of the $U$ and $V$ filters
are 3841 \AA, and 5479 \AA, respectively. Typical accuracy in the
rest-frame $U-V$ color is of 0.05 mag for $\sigma_z \sim 0.03 (1+z)$,
and 0.1 for $\sigma_z \sim 0.06 (1+z)$.

\section{Selection of  high-$z$ progenitors of early-type galaxies}
\label{sec:samples}

\subsection{The morphologically-selected sample}
\label{sec:ms}

The morphologically-selected sample was constructed by including all
galaxies classified by ZEST as Type~1, i.e., as early--type galaxies,
and all bulge-dominated Type~2.0 disk galaxies. The latter are located
in volumes of $PC$--space that are adjacent to the Type~1 early-type galaxies,
and are therefore characterized by very similar structural properties
(see Paper~I).  Indeed, Type~2.0 galaxies have bulge properties very
similar to the Type~1 systems, and differ from these only for a
clearly detected (non--dominant) disk component.  Down to the
magnitude limit of our analysis ($I_{AB}=24$), in the fraction of the
COSMOS field under investigation, and in the redshift range $0.2<z\le
1.0$ there are a total of 2352 and 1352 galaxies, with good
photometric redshift that are respectively classified as Type~1 and
Type~2.0 galaxies by ZEST. Both samples have similar Sersic $n$
distribution, with 80\% and 70\% of Type~1 early-type and Type~2.0 
bulge dominated galaxies,
respectively, having $n\ge 2.5$.  We highlight in the Sections below
further similarities between Type~1 and Type~2.0 galaxies.  The
inclusion of these (heavily) bulge-dominated disk galaxies in our
morphologically-selected sample of ETGs also facilitates the
comparison with other studies, since these have generally tended to
include such systems in the analysis \citep[see, e.g.,][]{bell2004b}.

\subsubsection{A further constraint: The z=0 Kormendy relation}
\label{sec:kormendy}

ETGs in the local universe follow well defined scaling
relations. Of particular importance is the fundamental plane
\citep{Djo1987,dressler} and, if the velocity dispersion is not
available, its photometric projection, i.e., the so-called
``Kormendy-relation'' \citep{kormendy1977}.  The Kormendy relation is
a correlation between the half--light radius of the galaxy and the
average surface brightness within the half--light radius.

Selecting ETG progenitors at high redshifts on the basis of their
elliptical-like morphologies is equivalent to requiring that these
progenitors are already fully assembled and dynamically-relaxed
galaxies at those earlier epochs: they become $z=0$ early-types by
evolution of their stellar populations (and possibly some modest
amount of additional star formation).  Therefore, in constructing a
morphologically-selected sample of high$-z$ ETGs, we can apply an
additional constraint, namely the surface brightness must be such that
with plausible surface brightness evolution it evolves in to the
Kormendy relation. Clearly, surface brightness evolution may be
uncertain, especially for blue objects with composite stellar
population, however we can take a conservative approach and estimate a
minimum evolution that leads to the minimum number of excluded
objects.

\begin{figure*}
\includegraphics[width=9cm]{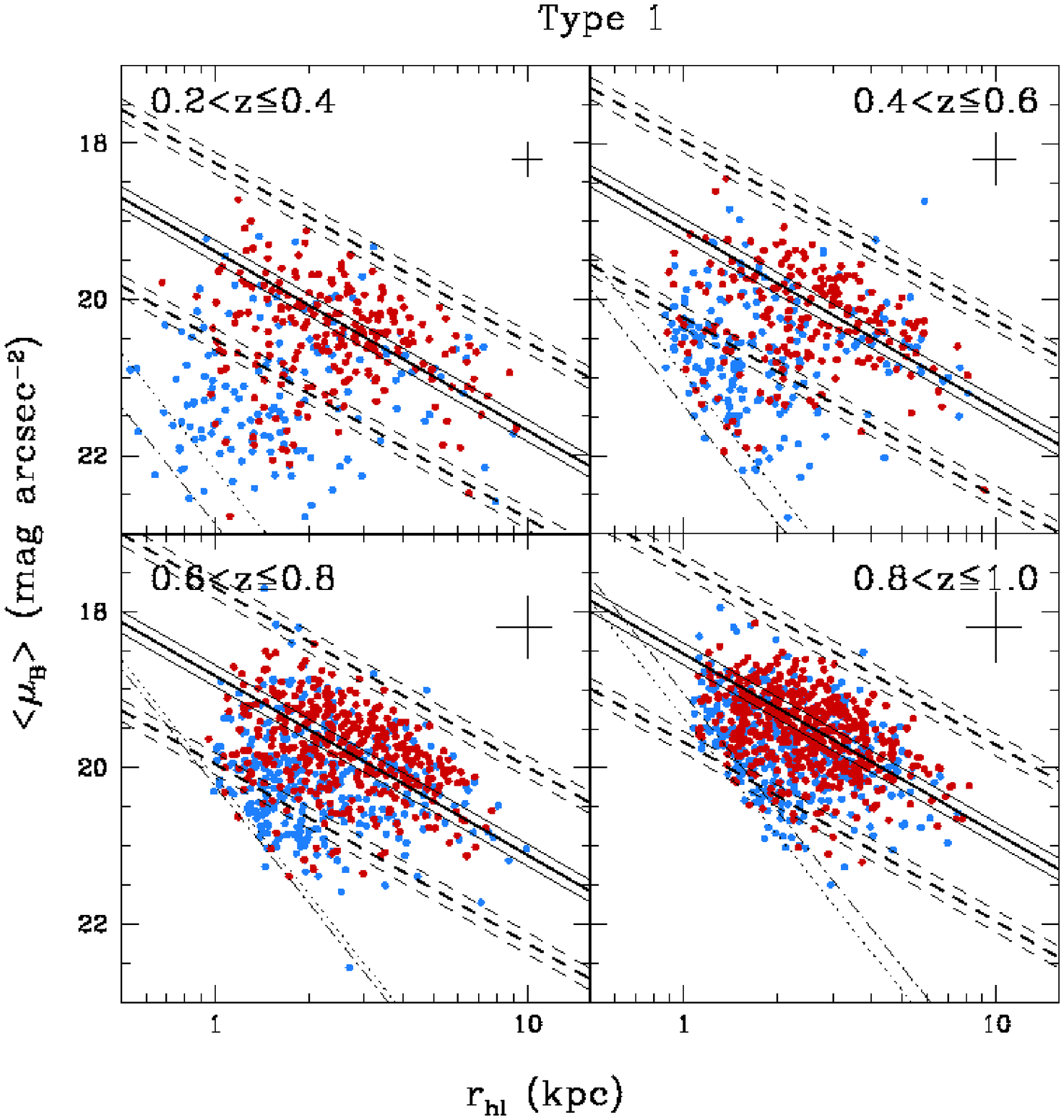}
\includegraphics[width=9cm]{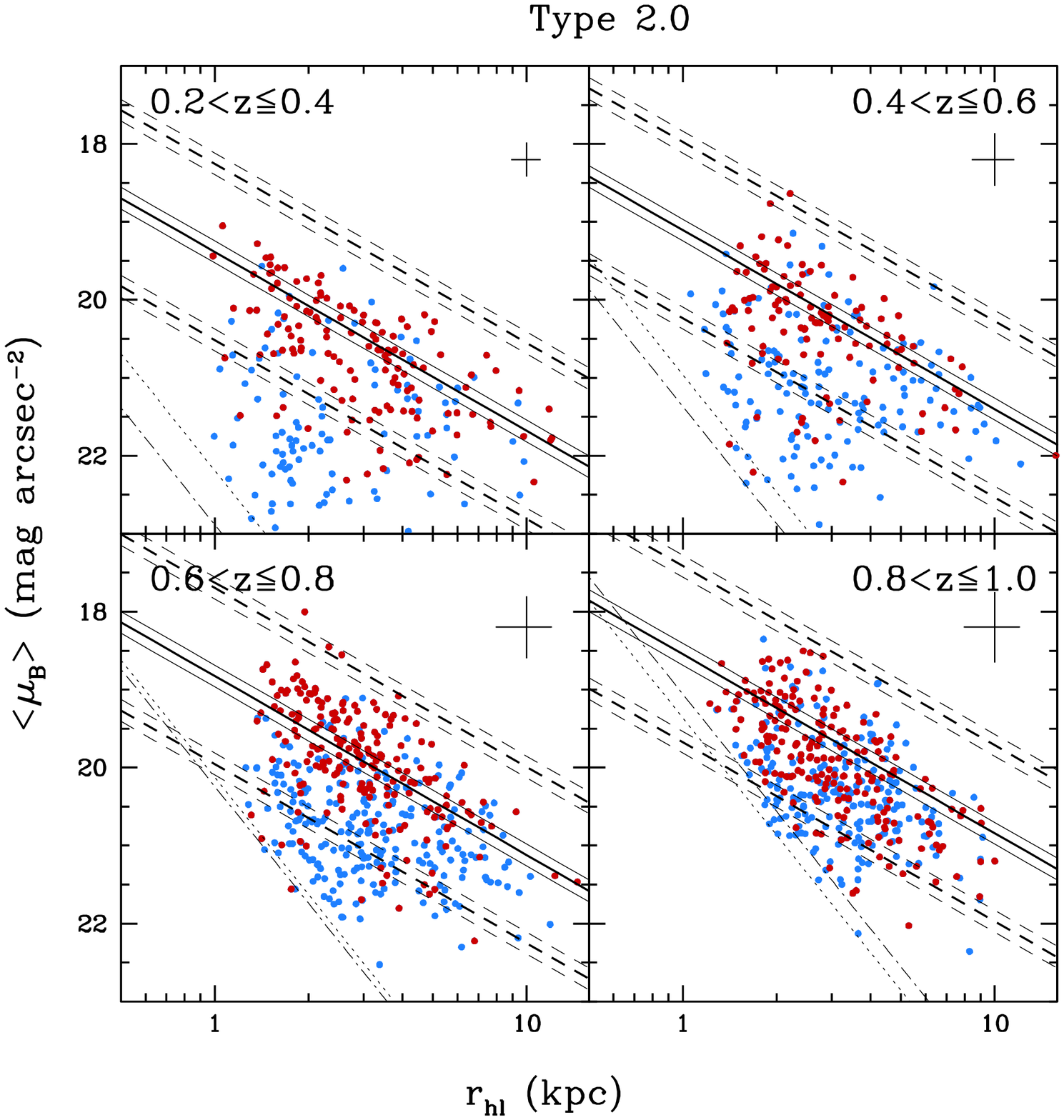}
\caption{Average surface brightness within the half--light radius in
  the rest-frame $B-$band ($\langle \mu_{B} \rangle$) as a function of
  the half--light radius ($r_{hl}$, in kpc) for all galaxies {\it
    morphologically-}classified as early-type galaxies (ZEST Type~1;
  left) and bulge-dominated disk galaxies (ZEST Type~2.0; right).  In
  each Figure, the four panels correspond to different redshift bins,
  as indicated.  Points are color--coded according to the best--fit
  ZEBRA spectral types: red points correspond to galaxies with ZEBRA
  elliptical--galaxy SED, and blue points represent all galaxies with
  later ZEBRA spectral types. The solid black line represents the best
  linear fit derived for the $z=0$ Coma cluster galaxies (J\o rgensen,
  Franx \& Kj\ae rgaard 1995), passively evolved at the central
  redshift of the considered bin. The solid thin lines represent the
  evolution of the best--fit relation at the extremes of the redshift
  bin. The dashed lines indicate a distance from the best--fit line of
  twice the ${\rm rms}$ of the $z=0$ relation.  Dotted and dot-dashed
  lines show, at the central redshift of bin, the $I_{AB}=24$
  magnitude limit for galaxies with the SED of a star-forming and an
  elliptical galaxy, respectively. On the upper right of each panel we
  show the typical error bar in $\langle \mu_B \rangle $ and
  $r_{hl}$.}
\label{fig:kormendy}
\end{figure*}

In Figure~\ref{fig:kormendy} we show the average surface brightness
within the half--light radius in the rest-frame $B-$band as a function
of the half--light radius for
the Type~1 (early-type galaxies, left panel) and Type~2.0 (bulge-dominated 
disk galaxies, right panel) galaxies of our
morphologically-selected sample. We use different colors to represent
galaxies with different ZEBRA best--fit to their SEDs; in particular,
red points are galaxies fitted by ZEBRA with an ``elliptical-galaxy''
template, and blue points represent later spectral types
\citep{feldmann2006}.  Dotted and dot-dashed lines show the
$I_{AB}=24$ magnitude limit at the central redshift of the bin, for
galaxies with the SED of a star-forming and an elliptical galaxy,
respectively.  Typical error-bars (i.e., estimated at the median size
and magnitude of the sample galaxies) are shown on the upper right
part of each panel. The errors are calculated by also taking into
account the effects of the photometric redshift. The half-light radii
used in Figure~\ref{fig:kormendy} are measured using a growth curve
analysis on the ACS $I_{AB}$ images.  By comparison with the
 half-light radii measured with GIM2D
by Sargent et al. (2007) we found that PSF effects; which are
accounted for with GIM2D, start to be important only for galaxies with
measured $r_{hl}<$0\farcs2. At these sizes, $r_{hl}$ is typically
overestimated by $\sim 25$\% compared with the GIM2D measurement.  We
find similar results by analyzing a set of $I_{AB}=22-24$ simulated
galaxies with Sersic index $n=2,$ and 4. We decided not to correct 
for this effect, since the correction depends on the exact shape of
the galaxy surface-brightness profile, which is unknown.
We note, however, that our sample includes
only $\sim 7$\% of galaxies with measured $r_{hl}<$0\farcs2, so the
impact of the PSF correction on the final selected sample of 
Kormendy-consistent galaxies is minimal.
In the Figure, the solid black line represents the best linear fit
derived for the Coma cluster galaxies \citep{jfk1995}, passively
evolved at the central redshift of the bin assuming $\mu_B(z) \propto
-1.36\,z$ [i.e., the evolution of a single stellar population model
with formation redshift $z_f=2$; \citet{bc03}].  The J\o rgensen et
al.  (1995) data were converted from Johnson VEGA to AB magnitudes
using the relation $B_{\rm AB}=B_{\rm VEGA} -0.1$. The solid thin
lines represent the best fit relation at the two extremes of the
redshift bins. The dashed lines indicate a distance from the $z=0$
best fit line of twice its ${\rm rms}$ dispersion.

Figure~\ref{fig:kormendy} shows that up to redshift $z\sim 1$, for a
given size, objects that have a ZEBRA elliptical-like classification
tend to have, on average, higher surface brightness relative to those
with later spectral types.  Furthermore, the Figure shows that, up to
$z\sim 0.6$, many galaxies lie significantly below the local Kormendy
relation.  A Fraction of order $\sim$80\% of these ``low-density''
systems have the ZEBRA SEDs of late-type, star-forming galaxies. We
note that at redshifts higher than $z=0.6$ selection effects are 
responsible for the absence of galaxies with $\langle \mu_B \rangle \sim 22$
mag arcsec$^{-2}$ and radii of few kiloparsecs.

In order to quantify the criterion for the exclusion of implausible
progenitors of ETGs on the basis of the $z=0$ Kormendy relation, we
followed the procedure described in \citet{ferreras2005}, namely:

\begin{itemize}

\item We assigned a fading rate of $\mu_B$ to each galaxy, based on
  the ZEBRA best fit to its SED.

\item We considered two extreme star formation histories (SFHs), both
  with an initial formation redshift $z_f=2$. The first SFH describes
  the passive evolution of a single burst stellar population and was
  assigned to all galaxies with a ZEBRA ``elliptical galaxy''
  best-fit.  The second SFH was assigned to galaxies best fitted by
  ZEBRA with a star-forming SED. This SFH is described by a constant
  star formation rate of 1M$_{\odot}/$yr up to the redshift at which
  the galaxy is observed (if larger than $z=0.5$), and by zero SFR
  from the galaxy redshift (or $z=0.5$) to $z=0$.  The lowest redshift
  possible for the truncation of the star formation, i.e., $z=0.5$,
  was chosen in order for the $z=0$ stellar population of the
  so-evolved galaxies to have colors consistent with the observed
  scatter in the color--magnitude relation of nearby ETGs
  \citep{bower1992,bernardi2003d}. The SFHs, i.e., the $\mu_B$ 
fading rates, for intermediate ZEBRA
  photometric types were derived by interpolating between the two
  extreme cases.  The evolution of the SFHs was computed assuming a
  constant metallicity equal to the solar value.
  
\item Finally, we excluded from the \underline{final} sample of
  ``Kormendy-compatible'', morphologically-selected progenitors of
  ETGs all objects that, when evolved to $z=0$ with the above
  evolutionary tracks, were found to lie away from the local Kormendy
  relation\footnote{The $z=0$ Kormendy-relation, as we discussed in
    the text, is derived using galaxies belonging to the Coma cluster.
    \citet{bernardi2003b} -see also discussion in \citet{renzini2006}-
    found that there are very small differences in the fundamental
    plane derived for galaxies in cluster or in the field, and these
    differences remain small up to redshift $\sim 1$
    \citep{vdkvdm2006}.}  at a distance more than $2$ times its
  observed scatter. Using a threshold of  $3$ times the observed
  scatter would increase the number of galaxies included in the final
  sample of Kormendy-compatible ETGs by $\lesssim 10$\%, without
  changing our main conclusions on the evolution of the number density
  of bright ETGs.

\end{itemize}
This selection might in principle be affected by the specific value of the 
adopted formation redshift ($z_f=2$) and the approach of using the ZEBRA
best fit templates to determine the $B-$band  surface brightness evolution 
of each galaxy. However, neither of these choices affects our 
final conclusions.

First, if some of the galaxies which have a ZEBRA elliptical-galaxy
fit were actually reddened star-forming objects, then the number 
of rejected objects would be  minimized. 

Assuming a slower $B-$band luminosity evolution (i.e., a higher
formation redshift of, e.g., $z_f = 3$) would change the expected
brightening at redshift $z=1$ by $\sim$0.2 magnitudes, resulting in a
final sample of kormendy-consistent ETGs only $\sim 7$\% larger than
the sample obtained using $z_f=2$.  Our choice of using $z_f=2$ is
supported by recent results, based on the evolution of the
fundamental-plane, showing that $M>10^{11}M_{\odot}$ early-type
galaxies have $z_f=2.01^{+0.22}_{-0.17}$ \citep{vdkvdm2006}, with no
significant differences for ETGs in cluster and field environment.
Althought $z_f$ measured by \citet{vdkvdm2006} is a luminosity weighted 
mean star-formation epoch, the star-formation timescale for massive ETGs,
estimated from $z=0$ ellipticals, is shorter than 1 Gyr.
Furthermore, \citet{dickinson2003} and \citet{papovich2006} show that
the star formation in massive galaxies is largely completed by $z\sim
1.5$. Although the evolution of the stellar-population for less massive
ETGs is less constrained, there is evidence that less massive ETGs
evolve faster than high mass ones, both in the field and in cluster
\citep{mcintosh2005,treu2005}. In the light of our results a mass
independent formation redshift is thus a conservative approach, since it 
gives the smallest number of rejected low luminosity ETGs.

The Kormendy-cut increases the fraction of galaxies in the
morphologically-selected (Type~$1+2.0$) sample which also have the
ZEBRA SED of an elliptical galaxy: this increases from 53$\%$ to
63$\%$.  Particularly affected are the lowest redshift bins up to
$z=0.6$, in which the fraction of early-type morphologies with an
early-type SED increases from 53$\%$ to 70$\%$.  We note that about
46\% of the excluded ZEBRA late-type galaxies at redshifts $z<0.6$ are
faint systems with $I_{AB}\ge 23$. In the highest redshift bin, the
fraction of excluded objects with ZEBRA late-type SEDs reduces to 20\%,
and only 1\% of the ZEBRA elliptical type objects do not pass the
Kormendy-relation test.

A similar fraction of Type 1 and Type 2.0 galaxies are excluded on the
basis of the Kormendy relation constraint: the fraction of Type~2.0
galaxies in the final {morphologically-selected sample} changes from
37\% to 35\% before and after the Kormendy-cut, i.e., it remains
basically constant.  The exclusion of the ``Kormendy-rejected''
objects implies a cut of about 20\% in the original sample of ZEST
morphologically-selected early--type galaxies: Our {\it \underline{
  final} sample of morphologically-selected} progenitors of ETGs
contains 2730 galaxies (1798 Type 1 and 932 Type 2.0 systems).

\subsection{The photometrically-selected sample}
\label{sec:cmselection}

Progenitors of $z=0$ massive ETGs may  appear, at $z\sim 1$,
as morphologically-irregular, merging systems with old and passively
evolving stellar populations.  These galaxies can be selected on the
basis of their SED properties.

To define our COSMOS sample of photometrically-selected progenitors 
of $z=0$ ETGs, we used
the red-sequence identified on the rest-frame $(U-V)$-$M_V$
color-magnitude diagram, as derived in the two-step procedure that we
describe below.

\begin{figure*}[ht]
\includegraphics[width=9cm]{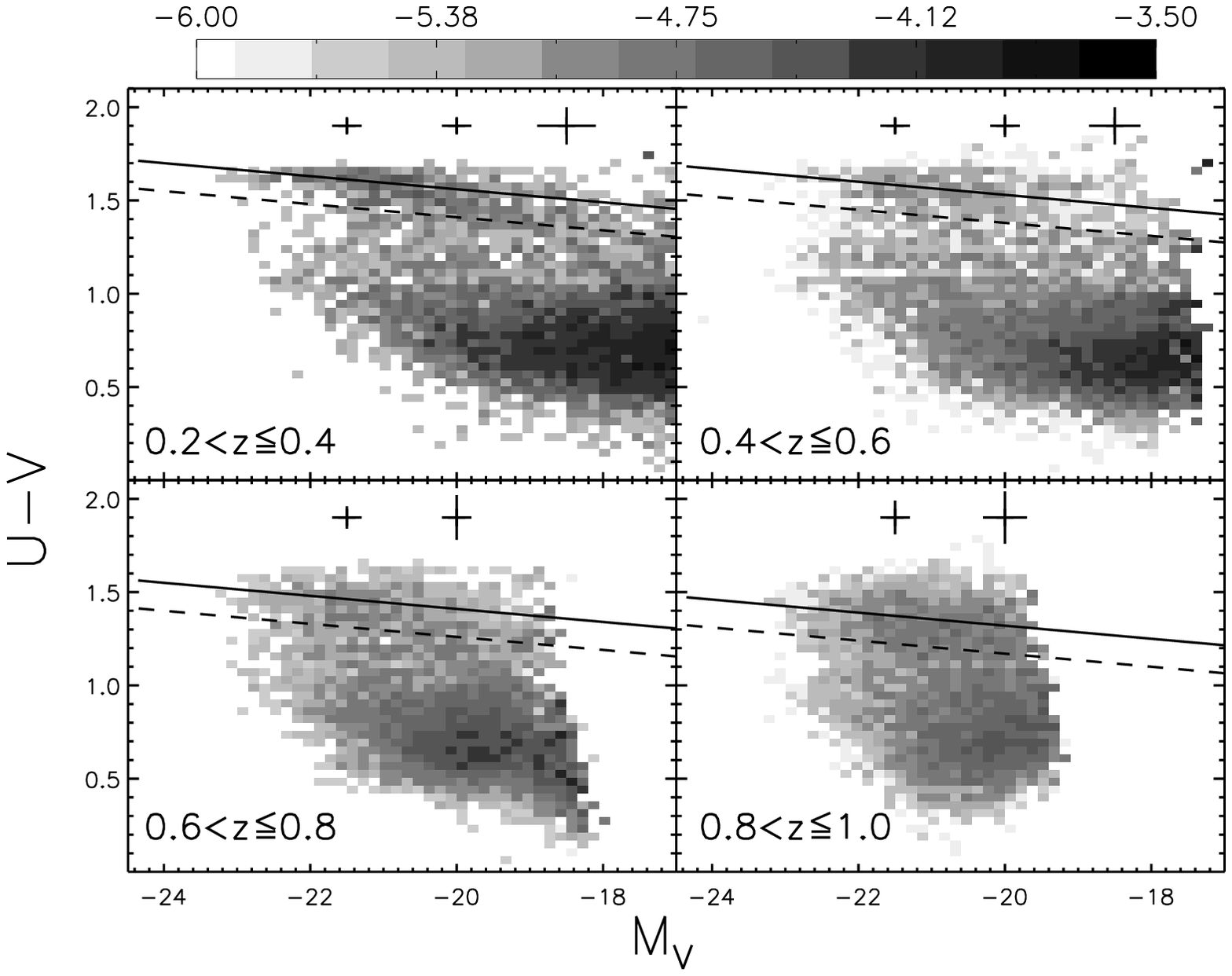}
\includegraphics[width=9cm]{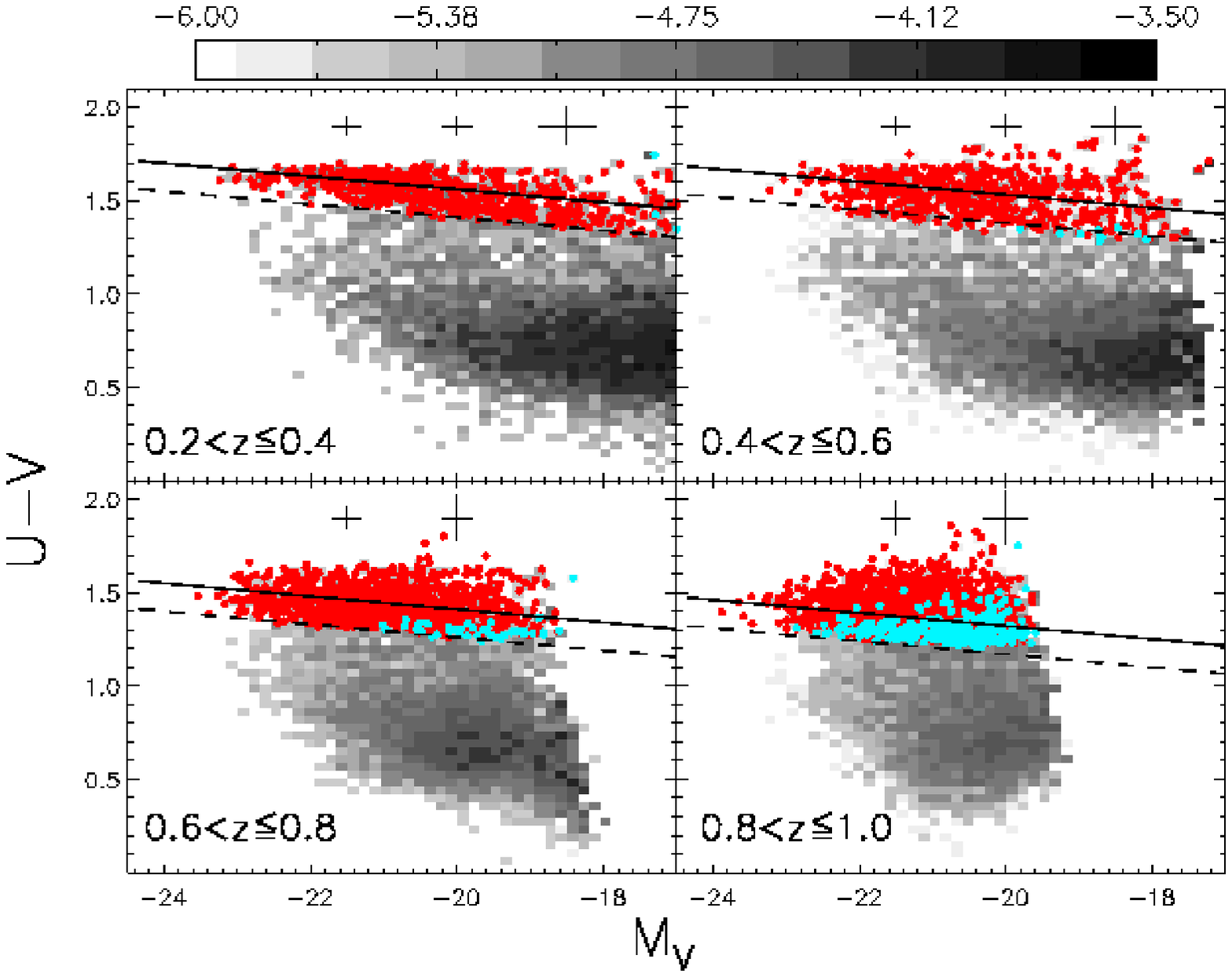}
\caption{Left panel: Rest-frame $U-V$ color versus absolute magnitude
  in the rest--frame $V-$band, for the same redshift bins as in
  Figure~\ref{fig:kormendy}.  In each redshift bin, the solid lines
  represent the color--magnitude relation of the red sequence at the
  central redshift of the bin (Section~\ref{sec:slope}); dashed lines
  are $-3\times {\rm rms}$ away from the solid lines. The grey levels
  represent the weighted volume density of galaxies in bins of 0.06 of
  $U-V$ color and 0.15 of magnitude, as indicated by the grey-scale
  shading bar given on top of each panel. Right panel: Same rest-frame
  color magnitude diagrams as in the left panel, with highlighted in
  red and blue the galaxies belonging to the photometrically-selected
  sample.  In this panel, red points indicate galaxies with a ZEBRA
  SED best fit of an elliptical-galaxy, and blue points are galaxies
  which belong to the photometrically-selected sample but have later
  spectral types according to the ZEBRA fits.}
\label{fig:CM}
\end{figure*}

\subsubsection{Step 1:  Identification of the "initial red sequence"}
\label{sec:slope}

In the left panel of Figure~\ref{fig:CM} we show the rest-frame
$(U-V)-M_V$ color--magnitude diagram for the COSMOS $I_{AB}\le24$
galaxy sample, split in four redshift bins. The gray levels in each
panel represent the weighted galaxy volume density in bins of 0.15 of
$M_V$ magnitude and 0.04 of $U-V$ color; the density in each
color--magnitude bin is computed as $\sum_i W_i/V_{max,i}$. The
$V_{max,i}$ value is the maximum volume within which a galaxy $i$,
with a given observed $I_{AB}$ magnitude and spectral type, is
detectable in the COSMOS survey; the weights $W_i$ are the corrections
required so as to account for objects excluded either because no
redshift is available, or because the redshift estimate is less
accurate than the required threshold (see Section~\ref{sec:photoz},
and also Paper~I for details on the derivation of the $W_i$ values).
The grey scale bar at the top of each panel shows the density in
Mpc$^{-3}$ corresponding to the grey intensity levels.

The presence of radial color gradients in galaxies can have an effect
on the measured colors and on the slope of the red sequence, depending
on the size of apertures used to compute colors and magnitudes
\citep[see, e.g.][]{scodeggio2001}.  Although not easy to quantify, an
estimate of this effect can be evaluated as follows. The galaxy colors
in the COSMOS photometric catalogue are computed within $3''$ diameter
apertures, regardless of galaxy redshift \citep{capak2006}. The local
magnitude-size relation implies a half--light radius of $ \sim 6$ kpc
at $M_B \sim -21.5$ and of $\sim 2$ kpc at $M_B\sim -19.5$; therefore,
when no evolution is taken into account, the COSMOS $3''$ aperture
correspond to $\sim 1.2$ and $\sim 3.4$ half--light radii for bright
and faint galaxies, respectively.  Assuming a De Vaucouleur profile
and using the average $U-V$ color gradient derived by Scodeggio (2001;
${\rm d}(U-V)/{\rm d}(\log{R})=-0.15$), the $U-V$ colors computed
within 1.5, 2.0, 2.5, and 5 $r_{hl}$ differ from the color computed
within 0.3 $r_{hl}$ respectively by $\sim -0.05$, $-0.07$, $-0.08$,
$-0.11$ magnitudes.  Given the assumptions on the shape of the galaxy
surface-brightness profile, on the value for the local color-gradient,
and on the size-luminosity relation, the estimates of the effect of
the color gradient on the measured colors can only be considered as
indicative, and we therefore decided not to correct the data, but
rather to consistently derive the slope of the red sequence from the
observed COSMOS color-magnitude diagram in each redshift bin.
       
\begin{figure}[ht]
\includegraphics[width=9cm]{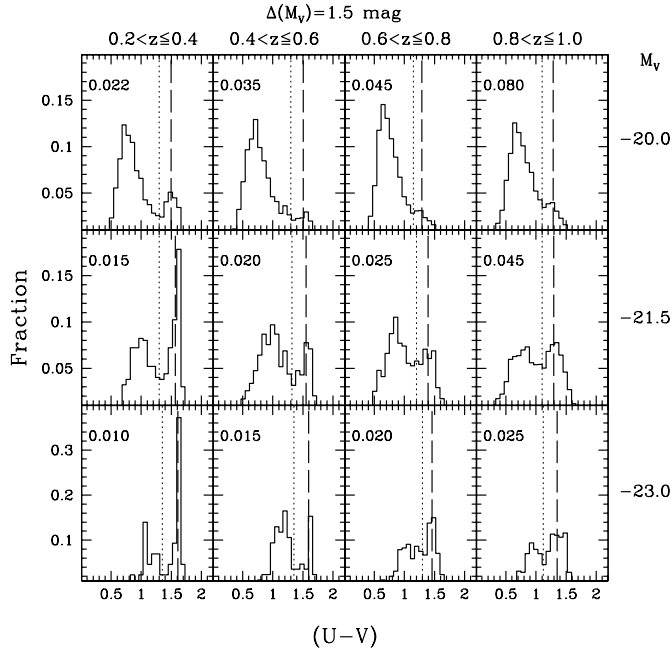}
\caption{Distribution of the rest-frame $U-V$ color for the COSMOS
  galaxies. We divide all galaxies in four redshift bins (one redshift
  per column) and three magnitude bins of 1.5 magnitudes for each
  redshift (as indicated on the right side of the plot).  The number
  in the upper-left corner of each panel indicates the median error on
  the $U-V$ colors in each magnitude--redshift bin
  (Section~\ref{sec:slope}). The dotted vertical line shows the
  ($U-V$)$_{irs}$ color that we use to broadly isolate the red
  sequence in each redshift-magnitude bin; the color of the peak in
  each bin (($U-V$)$_{redpeak}$) is indicated by the vertical dashed
  line.  The histograms are normalized to the total number of objects
  in each magnitude-redshift bin.}
\label{fig:newhistocolor}
\end{figure}

In Figure~\ref{fig:newhistocolor} we show the normalized distribution
of the rest-frame $U-V$ color of all galaxies split in both redshift
and absolute magnitude. In particular, we consider four redshift bins,
one per column, and three bins of 1.5 magnitudes centered at the
values indicated in the Figure (top: $M_V=-20$; middle: $M_V=-21.5$;
bottom: $M_V=-23.$).  All panels in the Figure show a degree of
bimodality between red and blue galaxies \citep[see e.g.,][for a
discussion of color bimodality at high redshift]{bell2004a}, albeit
with varying degrees of sharpness.  We are nonetheless able to
identify in each of the panels the color of the ``red peak'', which we
use as the starting point for the computation of the slope of the red
sequence in each of the individual redshift bins.

Specifically, in each magnitude--redshift bin, we define $(U-V)_{\it
  irs}$ as the color which broadly isolates the {\it initial red
  sequence}; $(U-V)_{\it irs}$ is defined as the color of the minimum
in the valley between the blue and red peak in each of the color
distributions, and it is indicated as a dotted line in each panel of
Figure~\ref{fig:newhistocolor}.  The $(U-V)_{redpeak}$ color is then
computed as the median of all colors redder than $(U-V)_{\it irs}$.
The $(U-V)_{redpeak}$ colors is indicated with a black dashed vertical
line in each panel of Figure~\ref{fig:newhistocolor}.  The number in
the upper-left corner of each panel indicates the median error on the
$(U-V)_{redpeak}$ colors in each magnitude--redshift bin; in some
bins, these are negligible.  These median errors are conservatively
obtained considering the maximum error derived by 
interpolating the errors in each of the two adjacent passbands 
that are used to derive the rest-frame
$(U-V)$ color of the galaxy. We tested that varying the adopted color
thresholds (i.e., $(U-V)_{redpeak}$) within $\pm 3 \sigma_{(U-V)}$
does not affect the resulting sample of photometrically selected ETGs.

We then compute the slope of the {\it initial red sequence} in each
redshift bin by fitting a linear relation to the $M_V -
(U-V)_{redpeak}$ points (and their errors).  The four slopes, as a
function of increasing redshift, are: $-0.037\pm 0.007$, $-0.032\pm
0.011$, $-0.040\pm 0.015$, and $-0.029\pm 0.020$. Within the errors,
these values are consistent with the average of $-0.035\pm 0.005$.
Given the measured uncertainties, we therefore choose to keep fixed in
our analysis the average slope of the red-sequence to the
``consistency value'' of $-0.035$($\pm0.005$) for all redshifts.

For each galaxy in each redshift bin, a ``slope-corrected'' color is
then obtained by setting $(U-V)_{corr} = (U-V)+ 0.035\, M_V$. The zero
point of the initial red sequence at the mean redshift of each bin is
thus defined as the color of the red peak of the $(U-V)_{corr}$
distributions.  These zero point colors $(U-V)_{ZP}$ are equal to
$0.86\pm 0.04$, $0.83 \pm 0.05$, $0.71 \pm 0.06$, and $0.62 \pm 0.05$
at the redshift values corresponding to the center of our bins, i.e.,
$z= 0.3, 0.5, 0.7$, and $0.9$, respectively. The error-bars are
computed by estimating the shifts in the red peak of the
$(U-V)_{corr}$ distributions induced by a change in the red-sequence
slope of $\pm 1 \sigma$.

The above-defined initial red-sequence curves are shown, for each
redshift bin, as solid lines in the left panel Figure~\ref{fig:CM}; in
each panel, the dashed line is $-3\sigma$ away from the initial red
sequence.

\subsubsection{Step 2: Refining the red-sequence: \\ The final photometrically-selected sample}
\label{sec:zp}

\begin{figure}[ht]
\includegraphics[width=9cm]{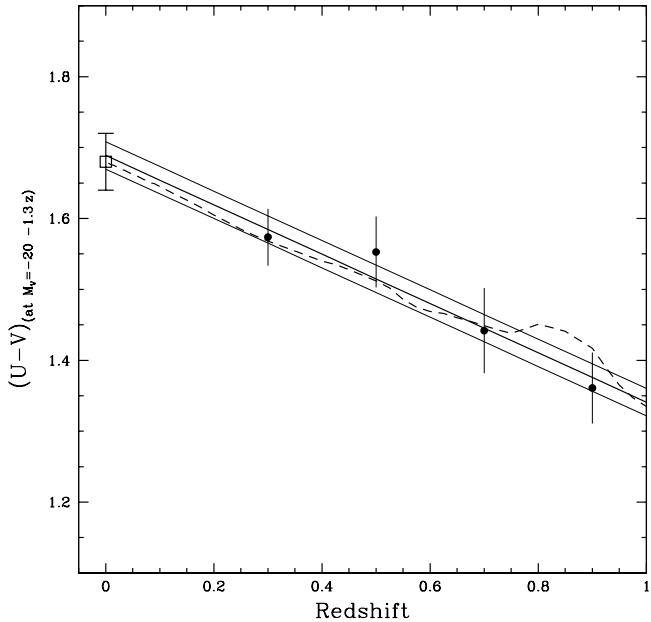}
\caption{The circles represent the $U-V$ color of the initial red
  sequence calculated, as a function of redshift, at a magnitude
  $M_V=-20.0$ passively evolved using a single-burst stellar
  population with formation redshift $z_f=2$ and metallicity
  $Z=0.8Z_{\odot}$. The color evolution of this model is shown as a
  dashed line. The error bars are those listed for the $(U-V)_{ZP}$ in
  Section~\ref{sec:zp}.  The $z=0$ point is derived applying the same
  procedure adopted for the COSMOS galaxies to the comparison SDSS
  sample.  The thick solid line shows the linear fit to the color
  evolution of the zero point of the red sequence, derived including
  the SDSS comparison data point. The thin solid lines are 1 $rms$
  away from the best fit.}
\label{figumb_z}
\end{figure}

The initial red sequence derived above coarsely mixes galaxies within
broad redshift bins.  To refine the evolution with redshift of the
zero point, we plot in Figure~\ref{figumb_z} the values of
$(U-V)_{ZP}$ derived above as a function of redshift.  Specifically,
Figure~\ref{figumb_z} shows the rest--frame $U-V$ color as a function
of redshift for an absolute $V$ magnitude of $-20.0$, passively
evolved by assuming a single burst stellar population formed at
redshift $z_f=2$ and metallicity $Z=0.8 Z_{\odot}$.  The color
evolution of the model is shown in Figure~\ref{figumb_z} as a dashed
line.

The $z=0$ point shown in Figure~\ref{figumb_z} is obtained by applying
to the comparison SDSS sample described in Appendix~\ref{app:sdss}
\citep[and discussed in detail in][]{kampczyk2006}, the same procedure
that we have described above for the COSMOS galaxies.  The observed
redshift evolution of the $U-V$ color of the red-sequence at $M_V=-20
-1.3\,z$ is consistent with the evolution predicted for the single
stellar population formed at redshift $z_f=2$, that  also reproduces
the $z=0$ $U-V$ color of the red-sequence of the SDSS ETG sample. 

A linear fit in the $0 \le z \le1$ redshift range (i.e., including the
SDSS point) to the $(U-V)_{ZP}$-$z$ relationship at the evolved
$M_V=-20$ magnitude gives:

\begin{equation}
 (U-V)_{ZP}=(-0.37 \pm 0.04)\, z + (1.69 \pm 0.02).
\end{equation}

This best fit is shown in Figure~\ref{figumb_z} as a thick solid line;
thin solid lines are located 1 $rms$ away from the best fit.  This
result is in agreement with the redshift evolution of ($U-V$)$_{ZP}$
found by \citet{bell2004a}.  We use this best fit, together with the
slopes measured as described in Section~\ref{sec:slope}, to define the
final red sequence at any redshift $z$.

We point out that the red sequence that is identified by our two-step
procedure is more robust towards contamination by interlopers than a
red sequence that is obtained, e.g., by keeping the slope fixed to a
value derived from local samples of ETGs \citep[e.g.,][]{bell2004a}.  Using a
non-optimally chosen slope introduces undesirable effects since this
basically determines the fraction of faint galaxies in the resulting
red sequence selected sample. Estimates for the slope of the $z=0$
red-sequence in local galaxy clusters range from $-0.12$ to $-0.02$,
depending on the aperture adopted for the measurements
\citep{scodeggio2001}. Assuming a too steep slope for the data under
study would lead to the inclusion of a large fraction of faint blue
interlopers.

The final red sequence defines our {\it photometrically-selected
  sample} of high-$z$ progenitors of $z=0$ ETGs. This sample includes
all galaxies with $U-V$ color redder than the value corresponding to
``minus $3\sigma$'' from the final red sequence at the appropriate
redshift, i.e., $U-V= 1.53 -0.37 z - 0.035\,( M_V+20) $.  The sample
contains 3844 galaxies with $I_{AB}\le 24$ and photometric redshift in
the range $0.2 <z \le 1.0$ \footnote{\citet[][]{bell2004a} found $\sim
  4700$ red-sequence selected galaxies in the redshift range
  $0.2<z<1.1$ down to $R_{\rm VEGA}=24$ in the COMBO17 survey, in an
  area similar to the COSMOS area presented here. If we consider the
  same redshift range, and relax our constrain on the accuracy of the
  photometric redshift we find $\sim 12$\% more
  photometrically-selected COSMOS ETGs than COMBO17. This is
  consistent with our being and $I_{AB}=24$ selected sample and their
  being a $R_{\rm VEGA}=24$ selected sample. We also stress that ours
  and Bell et al.'s samples are not directly comparable, since Bell et
  al. fixed the red-sequence slope to the value of $-0.08$ found in
  local galaxy clusters, while we consistently compute the red-sequence 
slope, from the observed color--magnitude diagram.}.

The right panel of Figure~\ref{fig:CM} highlights in color
the galaxies which belong to this photometrically-selected sample, and
specifically in red the galaxies that have a ZEBRA elliptical-galaxy
fit, and in blue those with ZEBRA late-type spectral fits.  Below
(Section~\ref{sec:kps}) we argue that it is sensible to apply the
Kormendy-test described above also to the so-derived
photometrically-selected sample; we will thus define as {\it
  \underline{final} photometrically-selected sample} those galaxies in
the sample constructed so far that also pass the "Kormendy-test".

\section{Properties of the morphologically- and photometrically-selected early-type galaxies}
\label{sec:results}

\subsection{Colors of morphologically-selected galaxies}
\label{sec:colors}

\begin{figure}[ht]
\includegraphics[width=9cm]{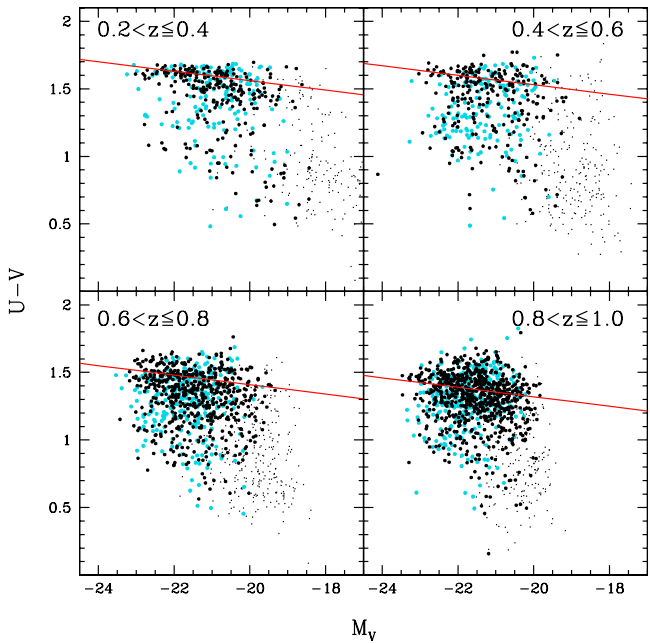}
\caption{Rest-frame color--magnitude diagram $U-V$ versus $M_V$ for
  the morphologically-selected sample of ETGs.  Black
  and cyan filled circles are used for galaxies kept in the sample,
  after passing the Kormendy-relation test described in
  Section~\ref{sec:kormendy}. Black filled circles are used for the
  ZEST Type~1 (early-type morphology) galaxies, while cyan filled
  circles are used for the ZEST Type~2.0 (bulge-dominated disk)
  galaxies. Black dots show the galaxies that were excluded from the
  morphologically-selected sample on the basis of the Kormendy test.
  The red line in each panel reproduces the final red-sequence derived
  in Section~\ref{sec:zp} calculated at the center of the redshift
  bin.}
\label{fig:colors}
\end{figure}

In Figure~\ref{fig:colors} we show, separately for the considered four
redshift bins, the $(U-V)$-$M_V$ color--magnitude diagram for the
Kormendy-test-consistent, morphologically-selected sample of
early--type galaxies discussed in Section~\ref{sec:ms}.  In
particular, black filled circles show the ZEST Type~1 early-type galaxies, and
cyan circles represent the bulge-dominated Type~2.0 disk galaxies.
The red line in each panel reproduces the final red-sequence
--calculated at the center of the redshift bin-- derived in
Section~\ref{sec:zp}.  At any redshift, the majority of
morphologically-selected early--type galaxies is characterized by red
colors (consistent with the final red sequence defined above);
however, a fraction of order 40\% of these systems has significantly
bluer colors than the red sequence (see Tables~\ref{tbl:samples} and
\ref{tbl:fraction}).  These galaxies would be missed in a pure
color-based selection. This fraction is similar to that of 
morphologically-selected galaxies at $z=0$ that do not belong to the 
red-sequence, see Table~1 in \citet{renzini2006}. 
The morphologically-selected sample of
ETG progenitors covers the entire observed range of rest--frame
$(U-V)$ colors, although very few of them  have
$U-V< 0.5$. This is not a consequence of the Kormendy-relation
selection that has been applied to the sample: this is demonstrated by
the black dots in each panel, which show the galaxies with
morphological Type~$1+2.0$ that are excluded from the final
morphological sample on the basis of the Kormendy-test.
These "Kormendy-test-rejected" galaxies are typically faint systems
(of order $M_V > -19.5$); there is however no obvious dependence of
this galaxy population on the $U-V$ color.

\begin{figure}[ht]
\plotone{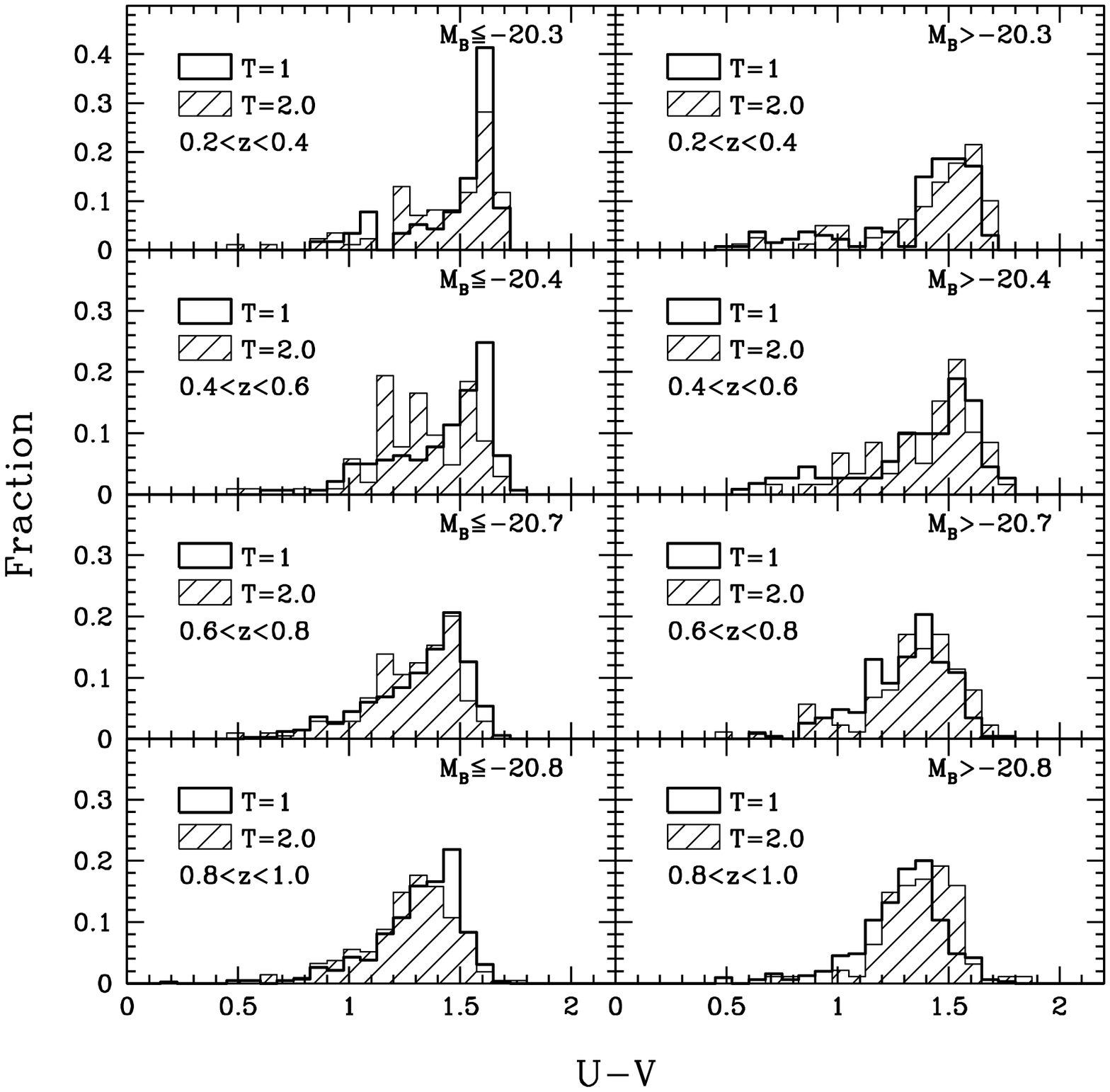}
\caption{The distribution of rest--frame $U-V$ color for the Type~1
  (early-type galaxies; solid histograms) and Type~2.0 
(disk galaxies; dashed histograms) galaxies that
  pass the Kormendy-relation test, and are thus kept in the final
  morphologically-selected sample of ETGs.  Left and right panels are for
  galaxies brighter and fainter than $M_B^*$, respectively.  The
  values of $M_B^*$ are derived from the Schechter fits to the LFs of
  the morphologically-selected sample (see text), and are given in the
  upper-right corners of the diagrams.}
\label{fig:histumvtypes}
\end{figure}

As evident from Figure~\ref{fig:colors}, the color distributions of
ZEST Type~1 and Type~2.0 galaxies are rather similar (apart from a
small blue ``excess'' at low redshifts of bulge-dominated disks at
bright magnitudes). This is further illustrated in
Figure~\ref{fig:histumvtypes}, where the rest--frame $U-V$ color
distributions are shown for both types as solid and dashed histograms,
respectively.  In each redshift bin, the Type~1 and Type~2.0 samples
are split in two components, one brighter (left panels) and the other
fainter (right panels) than the value of $M_B^*$ at the center of the
given redshift bin; the $M_B^*$ values are derived from the Schechter
fits discussed in Section~\ref{sec:lf}, and are given in the top-right
corner of each diagram.

\subsection{Morphology \& structure of photometrically-selected galaxies}
\label{sec:kps}

\begin{figure*}[ht]
\includegraphics[width=9cm]{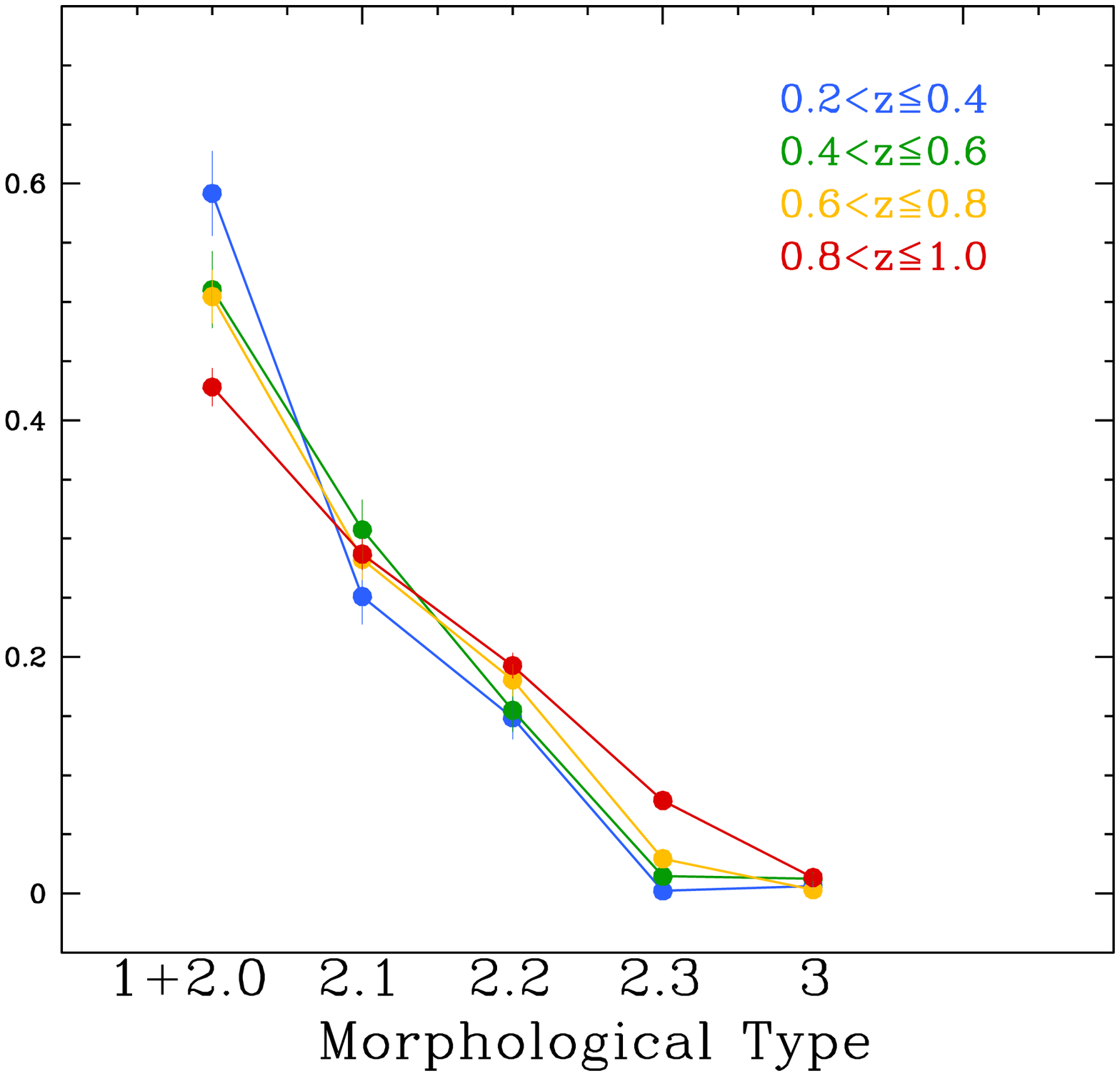}
\includegraphics[width=9cm]{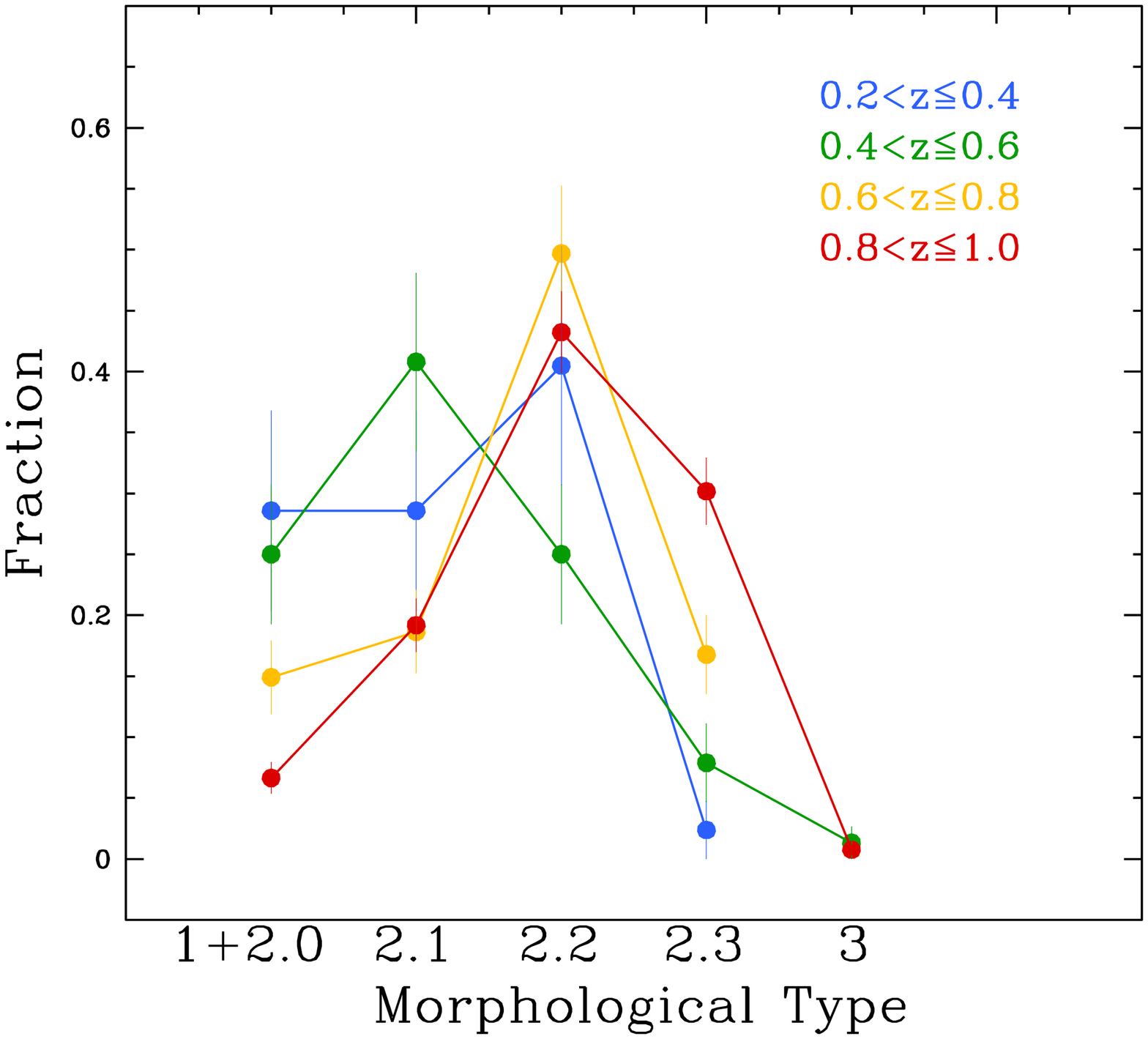}
\caption{Left panel: Fraction of photometrically-selected early-type
  galaxies with different ZEST morphological types (Type $1+2.0=$ 
morphological early-type; 2.1 to $2.3=$ disk dominated spirals; 
$3=$irregulars). In the right panel
  we show the results for the photometrically-selected galaxies which
  do not pass the Kormendy-test. To make the comparison among different 
redshift meaningful,  we consider (in both panels) only
  galaxies brighter than $M_B=-18.16, -18.44, -18.72$, and $-19$ at
  $z=0.3,0.5,0.7,$ and 0.9, respectively. Results for different
  redshift bins are shown in different colors.}
\label{fig:morphology}
\end{figure*}

In the left panel of Figure~\ref{fig:morphology} we show the fractions
of photometrically-selected early--type galaxies (derived in
Section~\ref{sec:cmselection}), that belong to different ZEST
morphological classes.  To make the comparison among different
redshift bins meaningful, we limit the analysis to $M_B=-19$ at
redshift $z=0.9$, and we evolve this magnitude limit as a function of
redshift according to the evolution of a single-burst stellar
population formed at $z_f=2$; according to the relation 
$\Delta M_{B}=-1.36z$ (c.f. Section\ref{sec:kormendy}).

In Figure~\ref{fig:morphology}, Type~1 early-type galaxies and 
Type~2.0 bulge-dominated galaxies are
considered as a single class (labelled with ``$1+2.0$'').  The results
for the four redshift bins are represented with different colors (as
described in the Figure). Galaxies with early-type morphologies, i.e.,
the combined sample of Type~1$+2.0$ galaxies, contribute from $\sim
45$\% at $z=0.9$ to $\sim$60\% at $z=0.3$ to the sample of
photometrically-selected ETGs. This compares to the fraction of $\sim$
58\% at $z=0$ \citep{renzini2006}. The majority of the galaxies that
are not classified as early--type Type~1 or Type~2.0 galaxies have the
morphology of disk-dominated, small-bulge galaxies ($\sim 30$\% and
$20$\% of ZEST Type~2.1 and 2.2, respectively); the remaining few
percents are almost equally split between bulgeless disks (Type~2.3)
and irregular galaxies (Type~3).

\citet{bell2004b} presented the morphological distribution of a sample
of COMBO17 red-sequence selected galaxies with redshifts between
$0.65\le z <0.75$. They find that, down to $M_V=-20.3$, 85\% of the
combined rest-frame $V-$band luminosity density comes from visually
classified E/S0/Sa galaxies.  A slightly smaller fraction (75\%) was
found by \citet{mcintosh2005} using an automatic morphological
classification for early--type galaxies based on the Sersic index $n$.  
When limiting the
analysis of the COSMOS data to the same bright cut in $M_V$ magnitude
and considering the same redshift range, we find that a similar fraction of $\sim
75$\% of the rest-frame $V-$band luminosity density comes from
Type~$1+2.0$ ETGs.

\subsubsection{Kormendy-test selection of the photometric sample}

In principle, mergers of old stellar sub-units ``caught in the act''
could have irregular morphologies and not satisfy the Kormendy-test
described above.  We nonetheless investigate which fraction of the
photometrically-selected sample is  excluded on the basis of this
test, and which are the properties of the rejected objects.
 
\begin{figure}[ht]
\includegraphics[width=9cm]{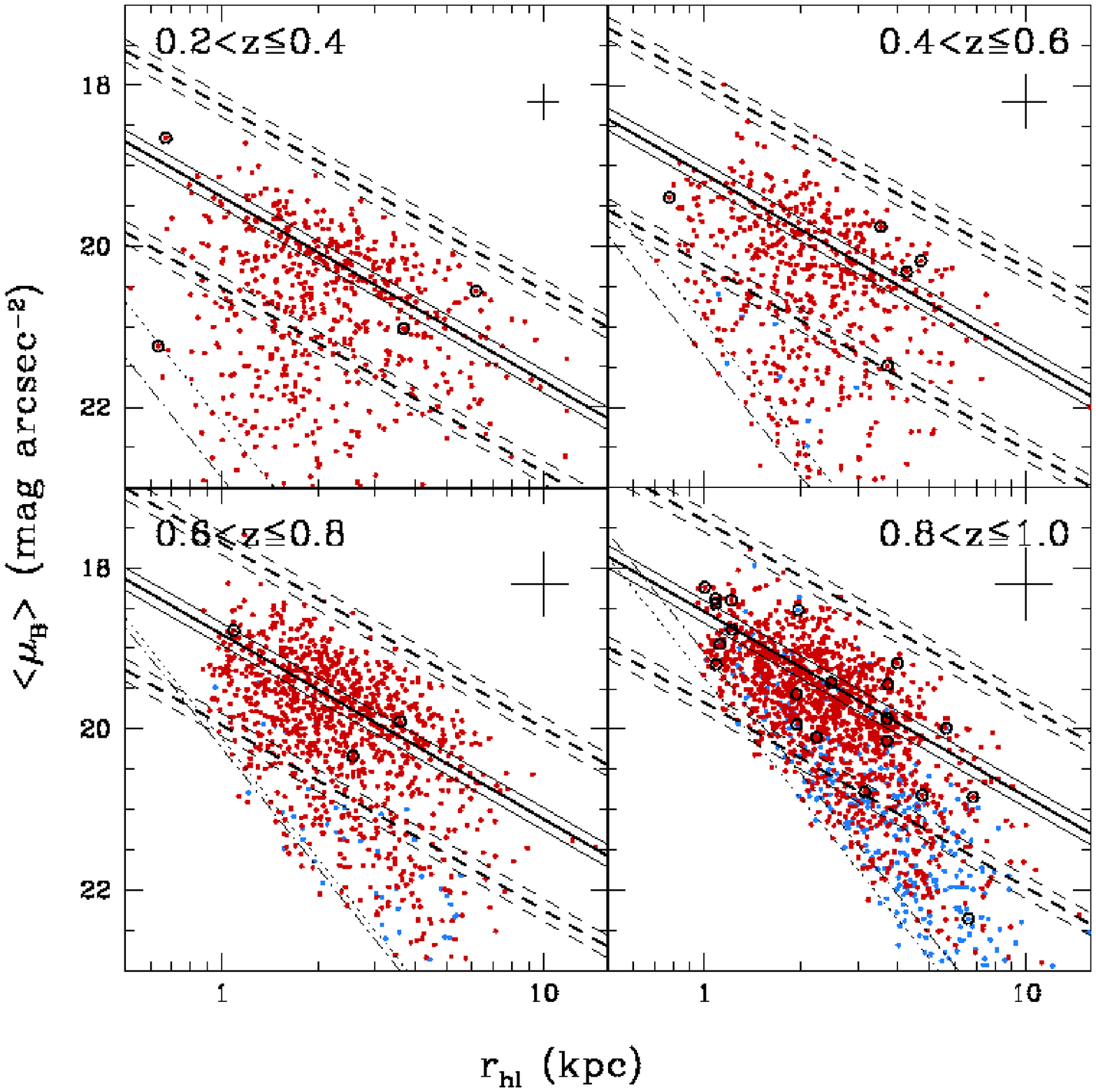}
\caption{The Kormendy diagram for the sample of
  photometrically-selected early--type galaxies of
  Section~\ref{sec:cmselection}. Large black circles identify galaxies
  with irregular morphological classification (ZEST Type~3).  Red dots
  are galaxies with SEDs best fitted by ZEBRA with an
  elliptical-galaxy template; blue dots identify objects which are
  fitted by ZEBRA with a late-type galaxy template.}
\label{fig:kormcolorselected}
\end{figure}

In Figure~\ref{fig:kormcolorselected} we show the Kormendy diagram for
the photometrically-selected sample of ETGs, split in the four
redshift bins; symbols and lines are as in Figure~\ref{fig:kormendy}.
In this Figure we also identify all galaxies that have an irregular
(ZEST Type~3) morphological classification by large black circles.
Indeed, not all of the red--sequence selected galaxies are consistent
with passively-evolving into the Kormendy relation at redshift $z=0$.
Interestingly, however, the vast majority of galaxies that are
excluded on the basis of the Kormendy criterion have a regular (ZEST
Type~1 or 2) morphology.  This is illustrated in the right panel of
Figure~\ref{fig:morphology}, where we show the morphological mixture
of photometrically-selected galaxies which do not pass the
Kormendy-test. Only up to $\sim$2\% of these Kormendy-test-rejected
galaxies have the Type~3 irregular morphology. This is consistent with
the rather short ($< 1$ Gyr) timescales during which major
dissipationless mergers appear as one irregular systems. At all
redshifts, the vast majority ($>$70\%) of Kormendy-test-rejected
objects in the photometrically-selected sample are disk-dominated
galaxies.  Therefore, when studying the redshift evolution of the LFs
of plausible ETG progenitors, it seems appropriate to apply the
Kormendy-test-rejection criterion also to the photometrically-selected
sample, as already done for the morphologically-selected sample. We
thus indicate with {\it \underline{final} photometrically-selected}
sample the 2877 galaxies, selected according to the red-sequence
criterion described in Section~\ref{sec:cmselection}, that furthermore
pass the Kormendy-test. We will nevertheless also show, in our
analysis of the LFs, the results obtained when no Kormendy-selection
is performed on the galaxy samples under study.

\subsection{The combined sample of progenitors of z=0 ellipticals}

There is, obviously, overlap between the morphologically-- and
photometrically--selected samples, as shown in
Figures~\ref{fig:colors} and Figure~\ref{fig:morphology}.  We
therefore also present, in the following, the analysis for the {\it
  combined sample} of all plausible candidate ETG progenitors at each
redshifts, i.e., {\it the union of the above-defined morphologically-
  and photometrically-selected samples}.  This combined sample
contains a total of 3980 galaxies; specifically, 1623 of the 2730
galaxies in the final morphologically-classified sample are also
members of the final photometrically-classified sample (see
Tables~\ref{tbl:samples} and \ref{tbl:fraction}).

\begin{center}
\begin{deluxetable}{cccc}
\tabletypesize{\scriptsize}
\tablecaption{Final samples of ETGs\label{tbl:samples}}
\tablewidth{0pt}
\tablehead{
\colhead{$z$(Range)} &
\colhead{MORPHO} &
\colhead{PHOTO} &
\colhead{COMBINED} \\
}
\startdata
 $0.2<z\le0.4$& 414  &  422 & 574    \\
$0.4<z\le0.6$ & 414  &  407 &594    \\  
$0.6<z\le0.8$ & 862  &  824 & 1214\\
$0.8<z\le1.0$ & 1040 & 1224 & 1598\\
\hline
\multicolumn{4}{c}{             }\\
TOTAL       & 2730& 2877       & 3980    \\
\enddata                           
\tablecomments{Number of galaxies in the final
  morphologically-selected (MORPHO), and photometrically-selected
  (PHOTO) samples, splitted in the four redshift bins. The number of
  galaxies per redshift bin in the combined sample are also listed.}
\end{deluxetable}
\end{center}

\begin{center}
\begin{deluxetable}{lcc}
\tabletypesize{\scriptsize}
\tablecaption{Overlap in the galaxy final samples\label{tbl:fraction}}
\tablewidth{0pt}
\tablehead{
\colhead{} &
\colhead{MORPHO } &
\colhead{PHOTO } \\
\colhead{$z$(Range)} &
\colhead{in PHOTO} &
\colhead{in MORPHO} \\
}
\startdata
$0.2<z\le0.4$ &63\%&61\% \\
$0.4<z\le0.6$ &54\%&55\% \\
$0.6<z\le0.8$ &55\%&57\% \\
$0.8<z\le1.0$ &65\%&54\% \\
\enddata                           
\tablecomments{ 
\\Listed are, split in the four redshift bins, the
  fraction of morphologically-selected galaxies present in the
  photometrically-selected sample, and the fraction of
  photometrically-selected galaxies present in the
  morphologically-selected sample.}
\end{deluxetable}
\end{center}

\subsection{Summary of the selected final samples}

For easy reference we summarize in Tables~\ref{tbl:samples} and
\ref{tbl:fraction}, as a function of redshift, the number of galaxies
contained in the morphological, photometric and combined final
samples, and the relative fraction of galaxies in common between the
morphological and photometric samples.

\section{The redshift evolution of the LF \\ of  early-type galaxies in COSMOS}
\label{sec:lf}

We derive the $B-$band LFs using the $1/V_{\rm max}$ estimator
\citep{schmidt1968,felten1976}: The number of galaxies per unit
comoving volume in the range of absolute magnitudes ${\rm d}M$, at
redshift $z$, and belonging to the sample $S$ can be written as:

\begin{equation}
\int{\Phi_S(M,z)}{\rm d}M =\sum_S{\frac{1}{V_{\rm max,i}}},
\label{eq:lf}
\end{equation}

\noindent
where the sum is over all galaxies in the sample $S$ within the
specific range of redshift and absolute magnitude.  $V_{\rm max,i}$ is
the maximum comoving volume within which a galaxy $i$ could still be
detected, given the apparent magnitude limits of the survey, i.e., in
our study, $16\le I_{AB}\le 24$; the volume is appropriately corrected
so as to account for the (small) fraction of objects that were
excluded from the initial ACS-selected sample due to the lack of
photometric redshifts.  More specifically, $V_{\rm max,i}$ is the
comoving volume between $z_1$ and $z_2$, computed, for each galaxy
$i$, taking into account the $k-$correction as a function of redshift.
The $z_1$ and $z_2$ values are, respectively, the maximum between
$z_L$ and $z_{16}$, and the minimum between $z_U$ and $z_{24}$; the
limiting values $z_U$ and $z_L$ are the upper and lower redshift of
the considered redshift bin, and $z_{24}$ and $z_{16}$ are the
redshifts at which the galaxy $i$ would have an $I_{AB}$ apparent
magnitude of 24 and 16, respectively, for a given rest--frame $B-$band
absolute magnitude and color.
  
The $1/V_{\rm max}$ estimator approach does not correct for sources
that are missed from the input catalog due to selection effects. In
particular, a catalogue that is defined according to a magnitude cut
in a given passband suffers from color--dependent selection effects
\citep[e.g.][]{lilly1995}. In our $I_{AB}\le24$ sample, blue objects
are detected down to fainter absolute $B$ magnitudes for redshifts
higher than $z\sim0.8$, i.e., the redshift at which the observer-frame
$I-$band coincides with the rest-frame $B-$band; below $z\sim0.8$, it
is the red objects that are instead detected to fainter absolute $B$
magnitudes.  Correcting for this bias is not straightforward, as it
requires an assumption on the color distribution of the galaxy
population that is missed from the sample.  To avoid this potential
source of error, we therefore limit the computation of the LF in each
redshift bin to the luminosity range for which we are complete,
regardless of galaxy colors.

\begin{figure}[ht]
\begin{center}
\includegraphics[width=9cm]{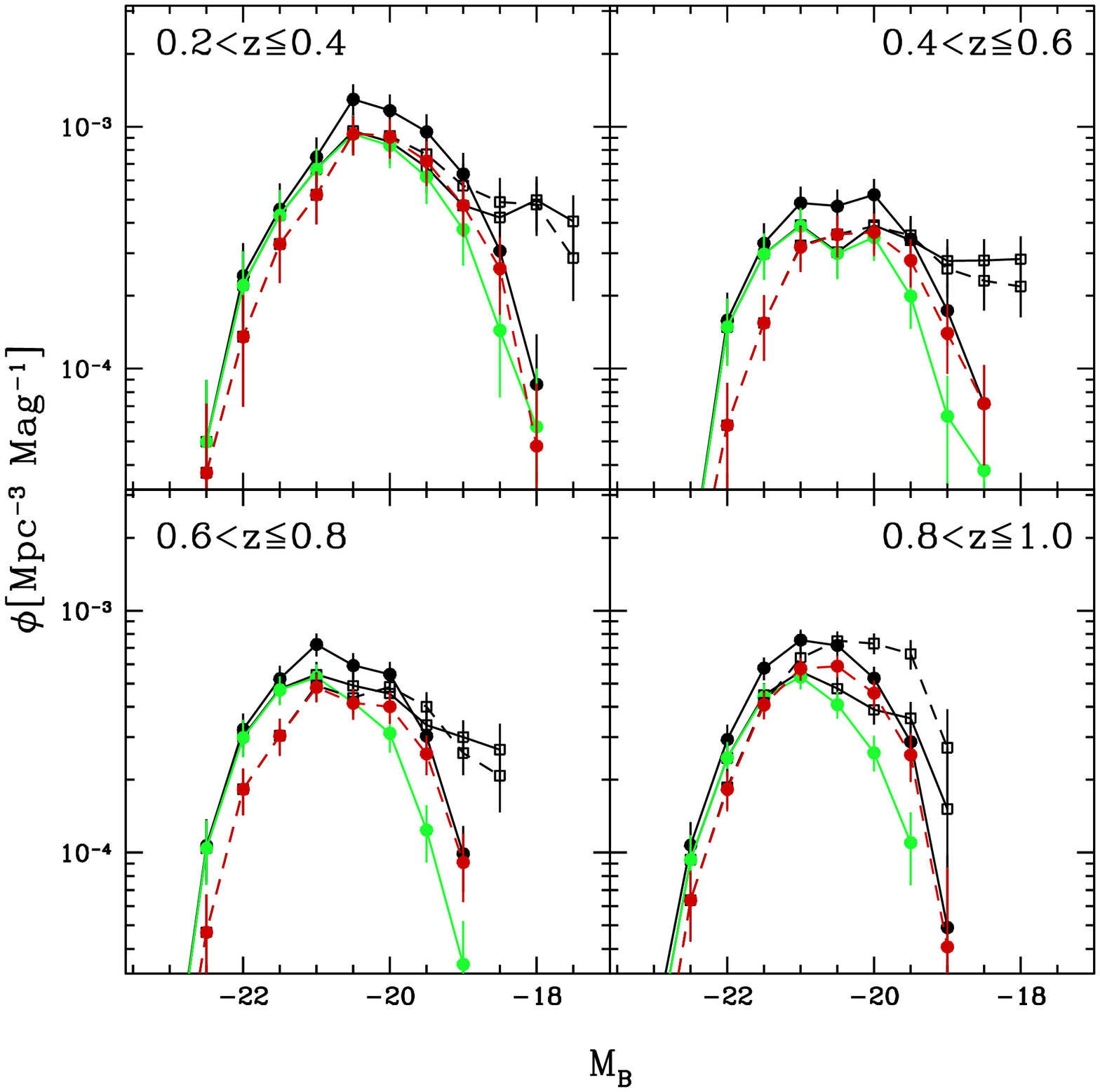}
\end{center}
\caption{LFs derived using the $1/V_{max}$ formalism and the ZEBRA
  Maximum Likelihood photometric redshifts. Green, red and black show
  respectively the morphologically-, photometrically-selected, and
  combined sample of COSMOS early-type galaxies.  The black open
  symbols and lines show the LFs derived without applying the Kormendy
  selection cuts (solid: morphological sample; dashed: photometric
  sample).  The error bars indicate Poisson errors only.  Note that
  the relatively high normalization factor in the lowest redshift bin
  is due to the well known overdensity in the central region of the
  COSMOS field. }
\label{fig:LF_ML}
\end{figure}

In Figure~\ref{fig:LF_ML} we show, split in four redshift bins, the
LFs computed for the different samples of ETGs described in the
previous Sections. Specifically, the green, red and black curves
represent respectively the {\it final morphologically-selected}
sample, the {\it final photometrically-selected} sample (i.e., both
after the Kormendy-test cut), and the {\it combined} sample of all
plausible progenitors of $z=0$ ETGs, which are selected either due to
their morphology or to their colors. The error bars in each luminosity
bin take into account Poissonian errors only.  
In Appensix~\ref{app:phzerr} we discuss all other sources of errors,
and their impact on the LFs. In particular, we show in the Appendix that 
our main conclusions are largely unaffected by uncertainties in the 
photometric redshift errors.

In the four panels of Figure~\ref{fig:LF_ML}, we highlight the effects
of cutting the sample according to the Kormendy-test. In particular we
show, with black empty symbols and thin lines, the LFs resulting from
the morphologically- (solid) and photometrically-selected (dashed)
samples when the kormendy-relation constraint is not applied. At all
redshifts, the inclusion of galaxies that are rejected on the basis of
the Kormendy-test would lead to a substantial increase in the number of 
ETGs at the faint end ($M_B>-20$) of the LFs.

Figure~\ref{fig:LF_ML} shows that at all redshifts, the shape of the
LFs of the final morphologically- and photometrically-selected samples
of ETGs are very similar.  However, at magnitudes brighter than
$M_B\sim-21$, morphologically-selected, i.e., dynamically-relaxed,
early--type galaxies tend to be more numerous than
photometrically-selected, i.e., old-stellar-population galaxies, by up
to a factor of $2$. In contrast, photometrically--selected ETGs are
more numerous relatively to the morphologically-selected by about the
same factor at fainter magnitudes. From the lowest up to the highest
redshift bin, respectively about $90 \pm 10$\% to $75\pm 10$\% of the
final sample of photometrically-selected ETGs brighter than
$M_B=-21-1.4\,z$ also belong to the morphologically-selected sample.

These results suggest that, at all redshifts under study, the majority
of the most massive ETGs are already dynamically-relaxed systems with
the appearance of $z=0$ elliptical galaxies and their stellar populations
are passively evolving, which is consistent with
other features of the LFs that we discuss in Section~\ref{sec:disc}.

\begin{center}
\begin{deluxetable}{cccc}
\tabletypesize{\scriptsize}
\tablecaption{Schechter Function best--fit parameters\label{tbl:schfit}}
\tablewidth{0pt}
\tablehead{
\colhead{$z$(Range)} &
\colhead{$\Phi^*$} &
\colhead{$M_B^*$} &
\colhead{$\alpha$} \\
\colhead{} &
\colhead{(Mpc$^-3$ Mag$^{-1}$)} &
\colhead{(mag)} &
\colhead{} \\
\colhead{(1)} &
\colhead{(2)} &
\colhead{(3)} &
\colhead{(4)} \\
}
\startdata
\multicolumn{4}{c}{Final morphologically-selected sample}\\
0.2--0.4 &$   0.0025\pm 0.0001$ & $ -20.34\pm   0.10 $ & $   0.28\pm  0.11$\\
0.4--0.6 &$   0.0010\pm 0.0001$ & $ -20.38\pm   0.12 $ & $   0.64\pm  0.12$\\
0.6--0.8 &$   0.0014\pm 0.0001$ & $ -20.70\pm   0.07 $ & $   0.60\pm  0.09$\\
0.8--1.0 &$   0.0013\pm 0.0001$ & $ -20.76\pm   0.12 $ & $   0.42\pm  0.13$\\
\multicolumn{4}{c}{Final photometrically-selected sample}\\
0.2--0.4 &$   0.0025\pm 0.0001$ & $ -20.22\pm   0.12 $ & $   0.12\pm  0.13$\\
0.4--0.6 &$   0.0010\pm 0.0001$ & $ -20.14\pm   0.10 $ & $   0.43\pm  0.13$\\
0.6--0.8 &$   0.0013\pm 0.0001$ & $ -20.55\pm   0.08 $ & $   0.38\pm  0.09$\\
0.8--1.0 &$   0.0017\pm 0.0001$ & $ -20.62\pm   0.07 $ & $   0.21\pm  0.10$\\
\multicolumn{4}{c}{Combined Sample}\\
0.2--0.4 &$   0.0034\pm 0.0001$ & $ -20.28\pm   0.10 $ & $   0.12\pm  0.11$\\
0.4--0.6 &$   0.0015\pm 0.0001$ & $ -20.37\pm   0.08 $ & $   0.40\pm  0.09$\\
0.6--0.8 &$   0.0019\pm 0.0001$ & $ -20.63\pm   0.06 $ & $   0.47\pm  0.07$\\
0.8--1.0 &$   0.0021\pm 0.0001$ & $ -20.73\pm   0.07 $ & $   0.23\pm  0.08$\\
\enddata 
\tablecomments{The columns are: (1) Redshift range; (2) Schechter
  function normalization $\Phi^*$, and error; (3) $M_{B}^*$, and
  error; (4) Slope of the faint end, $\alpha$, and error. The fits are
  performed on the {\it final} morphologically- and
  photometrically-selected samples, i.e., after the Kormendy-test cut,
  and on the combined sample derived from these final sub-samples. }
\end{deluxetable}
\end{center}

We fit a Schechter function \citep{schechter1976} to the COSMOS LFs,
to derive for them an analytical description.  The Schecter function
best--fit parameters, i.e., the characteristic magnitude $M^*_{B}$,
the volume density at $M^*_{B}$, $\Phi_{*}$, and the slope of the
faint-end, $\alpha$, are derived adopting a $\chi^2$ procedure. 
The best fit parameters for the Schechter
fits are shown in Table~\ref{tbl:schfit}.  The relatively high
normalization factor in the lowest redshift bin is due to the well
known overdensity in the central region of the COSMOS field (which is
also a reason for using the SDSS to determine the cosmic evolution of
the LFs in the $z=1$ to $z=0$ window).  In the fits, the faint-end
slope $\alpha$ is allowed to vary, and $\alpha$ values in the range
$\approx$0.1-0.6 are derived, with {\it formal} $1\sigma$ errors of
about $\sigma_{\alpha}\approx 0.1$; these values are different from 
the $\alpha \sim -0.6$ value adopted by, e.g., \citet{bell2004a}. However, 
these values of $\alpha$ are derived on the Kormendy-consistent samples, 
and are therefore not directly comparable with those used by other authors.
The faint end slope in the highest redshift bin is however not well
defined, given the magnitude cut of our sample; furthermore, despite
the variations in $\alpha$ values, all the measurements are consistent
within 3$\sigma$ with a constant faint end slope.  Thus, all in all,
the analytical fits quantitatively demonstrate the similarity,
discussed above, of the LFs as a function of redshift.

\subsection{The $z=0$ comparison sample}
The volume of Universe covered by the Cycle~12 COSMOS data in the
redshift range $0<z\le 0.1$ is only $\sim 5.5\times 10^3$ Mpc$^3$, and
includes less than 50 galaxies; therefore, cosmic variance and small
number statistics make this bin inadequate for comparisons with the
higher redshift bins to trace galaxy evolution down to $z=0$.
Furthermore, the $0.2 < z\le 0.4$ redshift bin is affected by a
relatively strong overdensity \citep{scoville2006b}.

Therefore, following the approach described in Paper~I, in order to
study the evolution of the LFs from $z\sim 1$ to $z=0$, we compare the
COSMOS LFs with a complete sample of galaxies extracted from the Sloan
Digital Sky Survey \citep[][]{york2000}, appropriately redshifted to
$z=0.7$ so as to provide a direct calibration point for the $0.6<z\le
0.8$ COSMOS data.  The generation of the redshifted SDSS images is
described in \cite{kampczyk2006}, and also summarized in Paper~I and
in Appendix~\ref{app:sdss}.  The SDSS redshifted images were analyzed
following the identical procedure that was adopted for the COSMOS
galaxies: $(i)$ The SDSS galaxies were morphologically classified with
ZEST; $(ii)$ The photometrically-selected sample of SDSS early-type
galaxies was determined according to the two-step procedure described
in Section~\ref{sec:cmselection}; $(iii)$ The Kormendy-cut was applied
to both the photometrical and morphological samples of ETG
progenitors; and $(iv)$ Finally, the LFs for the
morphologically-selected, photometrically-selected and combined
samples of redshifted SDSS ETGs were calculated, using the approach
described in Section~\ref{sec:lf}.  We stress that by applying exactly
the same procedure to the COSMOS and to the SDSS samples, the same
systematics apply to both the high-$z$ and the local comparison
sample.  Thus, the direct comparison of the two allows us to identify,
without biases, any evolutionary effect of the galaxy samples under
study from $z=0.7$ to $z=0$.

\section{Discussion}
\label{sec:disc}

\begin{figure}[ht]
\includegraphics[width=9cm]{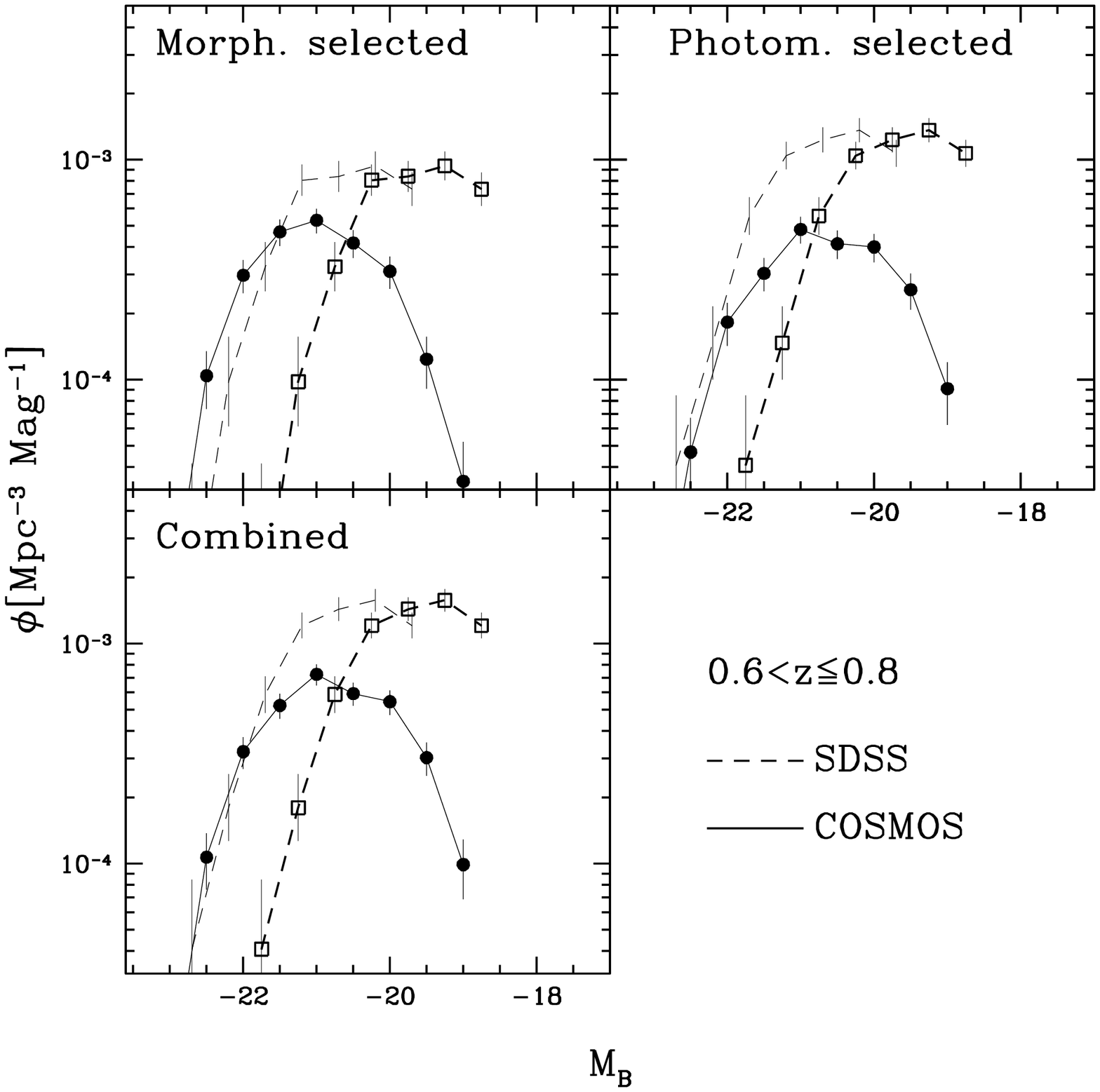}
\caption{Comparison of the SDSS LFs, redshifted to $z=0.7$ (empty
  symbols, dashed lines), and the $0.6<z\le 0.8$ COSMOS LFs (filled
  symbols, solid lines), for the morphologically-selected and
  photometrically-selected samples, and for the combined sample of
  ETGs.  The SDSS LFs are shown also brightened by $0.95$ magnitude,
  i.e., the amount consistent with the evolution of the red sequence
  in the rest-frame $U-V$ vs. $M_V$ color-magnitude diagram (thin
  dashed curves). }
\label{fig:lfsdss}
\end{figure}

In Figure~\ref{fig:lfsdss} we compare the $0.6<z\le0.8$ COSMOS LFs
(solid lines) with the redshifted SDSS LFs (thick dashed lines) for
the morphologically-selected, photometrically-selected and combined
samples of ETGs.  The thin dashed curves represent the SDSS LFs after
brightening by $\sim 0.95$ magnitude, i.e., the luminosity evolution
from $z=0.7$ to $z=0$ consistent with the observed evolution of the
red-sequence discussed in Section~\ref{sec:zp}.

At luminosities higher than $\sim 2.5-3\, L^*$ (i.e., $\sim 1$ mag
brighter than $M_{B}^*\approx -20.7$, see Table~\ref{tbl:schfit}), the
$z\sim0.7$ COSMOS LF of the photometrically-selected sample of ETGs
well matches the bright end of the corresponding SDSS LF, after
applying the brightening of $0.95$ magnitudes.  The
number density of the brightest --and thus most massive--
photometrically-selected ETGs does not evolve from redshift $z=0.7$ to
redshift $z=0$. At fainter magnitudes, however, the number density of
photometrically-selected early-type galaxies shows an increase of a
factor up to three from $z\sim0.7$ to $z=0$. We stress that the sample
of bright photometrically-selected ETGs is not
contaminated by dusty disks. Indeed, we found that the fraction of
Type~1$+$2.0 in the photometrically-selected sample of ETGs at
redshift $z=0.7$ is $\sim 90$\% for galaxies brighter than $M_B=-21.7$.

Broadly speaking, a similar result is obtained also for the
morphologically-selected and combined samples of early--type galaxies.
However, adopting a similar evolution as above, the number density of
the brightest, $M_B<-21.7$ morphologically-selected Type~$1+2.0$ ETGs
at $z\sim0.7$ in the COSMOS field would slightly exceed the number
density of the brightened SDSS sample.  This suggests a faster
luminosity evolution for the morphologically-selected sample of ETGs,
which is consistent with the latter containing a fraction of galaxies
with colors bluer than those predicted for a passively evolving
stellar population (see Section~\ref{sec:colors}).  These blue
colors might  arise from a substantial fraction of mass in
intermediate-age stars. However, a recent second bursts of star formation,
producing only a few percent of the stellar mass, would also make
significantly bluer the colors of the stellar population and increase
the $B-$band luminosity of the galaxies.

For example, adding a 1 to 5\% stellar mass in a second burst of star
formation to an underlying old, single-burst population with formation
redshift $z_f=2$ is sufficient to explain the $U-V$ colors of even the
bluest morphologically-selected ETGs at all redshifts under study.
This suggests that the bulk of the stellar mass in {\it all}
morphologically-selected high-$z$ ETGs is already formed by $z\sim 1$.
This is consistent with the other studies that we have already
mentioned, which find that $z\sim 0.3-0.75$ galaxies with $M \ge
10^{11} M_\odot$ have an integrated specific star formation rate less
than the global value. These studies conclude that the bulk of star
formation in massive galaxies occurs at early cosmic epochs and is
largely complete by $z\sim1.5$, and that further mass assembly in
these galaxies takes place with low specific star formation rates
\citep[e.g.,][]{dickinson2003,papovich2006}.

Furthermore, similar to the case of the photometrically-selected
progenitors of ETGs, also the morphologically-selected sample and the
combined sample of early--type galaxy progenitors show a remarkable
lack of evolution of the very bright-end of their LF. Fainter
early--type galaxies are instead again about a factor $\sim 2-3$ less
numerous than at $z=0$.

More quantitatively, we compute the total density of ETGs brighter
than $2.5\,L^*$ in both the combined sample of COSMOS and SDSS ETGs.
We indicate these two densities with $\rho_{COSMOS}$ and
$\rho_{SDSS}$, respectively. The adopted value of $L^*$ is the one
derived from the Schechter fit to the $z=0.7$ LF of the combined sample
(see Table~\ref{tbl:schfit}).  In order to compute $\rho_{SDSS}$, we
integrate the SDSS local LF brightened by 0.95 magnitudes.  Consistent
with the qualitative analysis presented above, we find that

\begin{equation}
  \rho_{COSMOS}/ \rho_{SDSS}=\frac{(2.32\pm 0.17)\times 10^{-4}}{(2.38\pm 0.20)\times 10^{-4}}=0.97\pm 0.11.
\end{equation}

\noindent
The density of $L>2.5L^*$ ETGs remains constant, within the Poissonian
errors, since $z=0.7$ to today.  This conclusion might in principle
depend on the assumed amount of luminosity evolution (i.e., the SFH of
the ETG population), on the cosmic variance, and 
on the effect of the photometric redshift
uncertainties on the LFs  As we discuss in
Appendix~\ref{app:phzerr} and Section~\ref{sec:lf}, Poissonian errors
at such bright magnitudes are, however, larger than the uncertainties
on the LF induced by the errors on the photometric redshifts. Errors
due to the photo$-z$ can thus be neglected.

The effect of cosmic variance on $\rho_{COSMOS}$ can be quantified
using the prescription of \citet{somerville2004}. We estimate that,
for the given density and COSMOS volume, this uncertainty contributes
at most $\sim$25\% to the error on $\rho_{COSMOS}$. We obtain a
similar result by calculating the variance of the total number of
$M_B<-21.7$, i.e. $>2.5L^*$, galaxies in the 24 mock galaxy catalogs
generated by \citet{cosmosmock} for the COSMOS survey using the
Millennium Run numerical simulations. The SDSS density, $\rho_{SDSS}$,
is not affected by cosmic variance, since the 1800 SDSS galaxies were
chosen randomly in a local volume larger than $2\times 10^5$ Mpc$^3$,
and the derived SDSS LF closely matches the \citet{blanton2003} LF
derived from the entire SDSS galaxy catalog \citep[see further
discussion in ][]{kampczyk2006}.

What causes the largest uncertainty in $\rho_{SDSS}$ is the assumed
luminosity evolution for the ETG stellar population. In the previous
analysis we have used the evolution of a single stellar population
formed at redshift $z_f=2$, justified by the fact that this SFH is
consistent with the observed evolution of the red-sequence zero point
(see Section~\ref{sec:zp}). However, we can estimate the changes
induced on $\rho_{SDSS}$ due to a reasonable range of SFHs, and use 
them as more realistic estimates of the errors on $\rho_{SDSS}$. Although
the range of observed $d(M/L)/dz$ reported in the literature is broad
\citep[see, e.g., Table~2 in ][and references therein]{treu2005}, a
recent study by \citet{vdkvdm2006} shows that stars in massive ETGs
($M>10^{11}M_{\odot}$), both in clusters and in the field, have a
mean luminosity-weighted formation redshift of
$z_f=2.01^{+0.22}_{-0.17}$.  For such massive galaxies, 
\citet{thomas2005} show that the star-formation time scale is less
than $1$ Gyr, which therefore makes our $z_f$ a reasonable approximation 
for the measured value. This formation timescale, together with the error on
$z_f$ presented by \citet{vdkvdm2006}, can be used to derive the
uncertainty on $\rho_{SDSS}$ due to varying the luminosity evolution
of the ETGs. We find that an uncertainty in formation redshift of
$\Delta z_f =0.4$ corresponds, roughly, to a $\Delta(M_B)=0.1$ at a
redshift $z=0.7$.  It follows that the total density of $M_B=-21.7 \pm
0.1$ ETGs in the combined sample is $\rho_{SDSS}=(2.38\pm 0.42)\times
10^{-4}$ Mpc$^{-3}$. It is clear, then, that the largest source of
uncertainty on $\rho_{SDSS}$ is indeed due to the assumed SFH.

Considering together the effects of cosmic variance and varying the
SFHs, the ratio between the observed and predicted density of massive
ETGs at redshift $z=0.7$ is therefore $\rho_{COSMOS}/\rho_{SDSS}=0.97
\pm 0.32$.  This shows that bright ($L>2.5\,L_*$) ETGs are already in
place at $z=0.7$, and that the maximum evolution in the number of
bright ETGs allowed from $z=0.7$ to $z=0$ is at most of $\sim 30$\%.  

In contrast, there is indeed a dearth of lower-luminosity ETGs, at
$z\sim 0.7$ compared with the local universe. This deficiency of
intermediate-to-faint ETGs is not an effect of incompleteness, since
($i$) both in the morphologically- and photometrically-selected
samples, the deficit of early--type galaxies is visible at magnitudes
$M_B\sim -21$, i.e., well above our magnitude limit ($M_B\sim -18.5$ at
$z=0.7$); ($ii$) the central wavelength of F814W filter matches the
rest-frame $B-$band at $z=0.8$; this implies that, down to the
faintest magnitude bin considered, there are no color-dependent
selection effects in the redshift range $0.6<z<0.8$; ($iii$) the ZEST
classification recovers the morphological class for $\sim 80$\% of the
galaxies down to $I_{AB}=23$ (Paper~I); furthermore, possible systematics
in the morphological classification are identical for the COSMOS and
for the local SDSS sample used to normalize our results.

Arguments for an apparent anti-hierarchical mass assembly of ETGs were
recently presented based on the redshift evolution of the mass
function \citep{bundy2005} and of the luminosity function (Cimatti,
Daddi \& Renzini 2006, Brown et al. 2007, Wake et al. 2006).
Furthermore, enough star-formation is observed in massive galaxies at
higher redshifts to account for all stars that we observe already
assembled in massive galaxies by $z\sim 1$ \citep{daddi2005}. Our
morphological selection of ETGs adds to this picture the key
information that not only the full mass assembly of the most massive
ETGs was completed at an earlier cosmic time compared to the less
massive galaxies, but also that the majority of these systems achieved
dynamical relaxation by $z\sim 1$.  A relatively large fraction of the
less massive early-type galaxies has either not yet achieved
relaxation at the redshift (epoch) at which they are observed, or not
completed its star-formation.

\section{Concluding remarks}

We highlight our three main results: $(1.)$ The shapes of the LFs 
of morphological, photometrical and combined samples of ETG
 progenitors at all redshifts are remarkably
similar; $(2.)$ The vast majority of photometrically-selected
massive ETGs is already dynamically-relaxed --and has thus the
morphology of an early-type galaxy-- at redshifts $z\sim 1$; and 
$(3.)$ There is a a deficiency of approximately $<2L^*$ ETGs as
opposed to remarkable constancy of the bright end ($L>2.5L^*$) of the
LFs of morphologically- and photometrically-defined ETGs.

These findings support a scenario in which the brightest, most massive
early type galaxies are already fully assembled $\sim 8$ Gyrs ago,
while fainter, less massive early--type galaxies keep forming their
stars and assembling their mass from $z=0.7$ to the present.  This
trend in the observed evolution of the LFs of early-type galaxies,
together with the observed constancy with redshift of the LFs' shapes,
argues against a significant contribution of dry mergers in the build
up of the $z=0$ massive ETG population. Furthermore, we suggest that
the $z=0$ lower luminosity early-type galaxies are the end product of
the conversion of blue irregular and disk galaxies into `red'',
early-type galaxies. This explanation is also supported by the
observed increase of about a factor of three in the number density of
irregular galaxies from the local Universe to $z=0.7$ (Paper~I).

\appendix
\section{Impact of photo-$z$ uncertainties on LFs}
\label{app:phzerr}
Feldmann et al. (2006) estimated the quality of the photometric
redshifts by comparison with spectroscopic redshifts for a sample of
$866$ galaxies observed within the $z-$COSMOS survey
\citep{lilly2006}.  The comparison shows that the ZEBRA photometric
redsfhits have an accuracy of $\sigma_z \propto 0.027\,(1+z)$ for
galaxies brighter than $I_{AB}=22.5$, and with redshift $z\le 1.2$.
In order to estimate the accuracy at magnitudes fainter than
$I_{AB}=22.5$, we simulated a faint version of the spectroscopic
redshift catalog by dimming the $I_{AB}\le 22.5$ galaxies with
available spectroscopic redshift down to our magnitude limit of
$I_{AB}=24$. We then recomputed the ZEBRA photometric redshifts for
these artificially-fainted sources.  The accuracy of the ZEBRA
photo$-z$'s down to $I_{AB}=24$ is $\sigma/(1+z)\sim 0.06$.

To estimate the effect of the photometric uncertainties on the
calculation of the LFs, we create one hundred versions of the COSMOS
catalog, by convolving the measured photometric redshift with the
photometric redshift error appropriate for the magnitude of each
object.  We consider two bins of magnitude: for all galaxies with
$I_{AB} \le 22.5$ we used $\sigma_z=0.03(1+z)$, while we use
$\sigma_z=0.06(1+z)$ for all galaxies with $22.5<I_{AB}\le 24.0$.
Although the photometric redshift errors should be applied to 
an ``error--free'' catalog, our test provides a conservative estimates
of how important the effects of the photometric redshift errors 
are on the observed LFs.

Using the simulated COSMOS catalogs we extracted 100 versions of the
morphologically- and photometrically selected samples of ETGs,
following the procedures described in Sections~\ref{sec:ms} and
~\ref{sec:cmselection}. We computed the LFs for these 100 re-simulated
samples using the procedure described in Section~\ref{sec:lf}.

\begin{figure*}[ht]
\epsscale{1.} 
\plotone{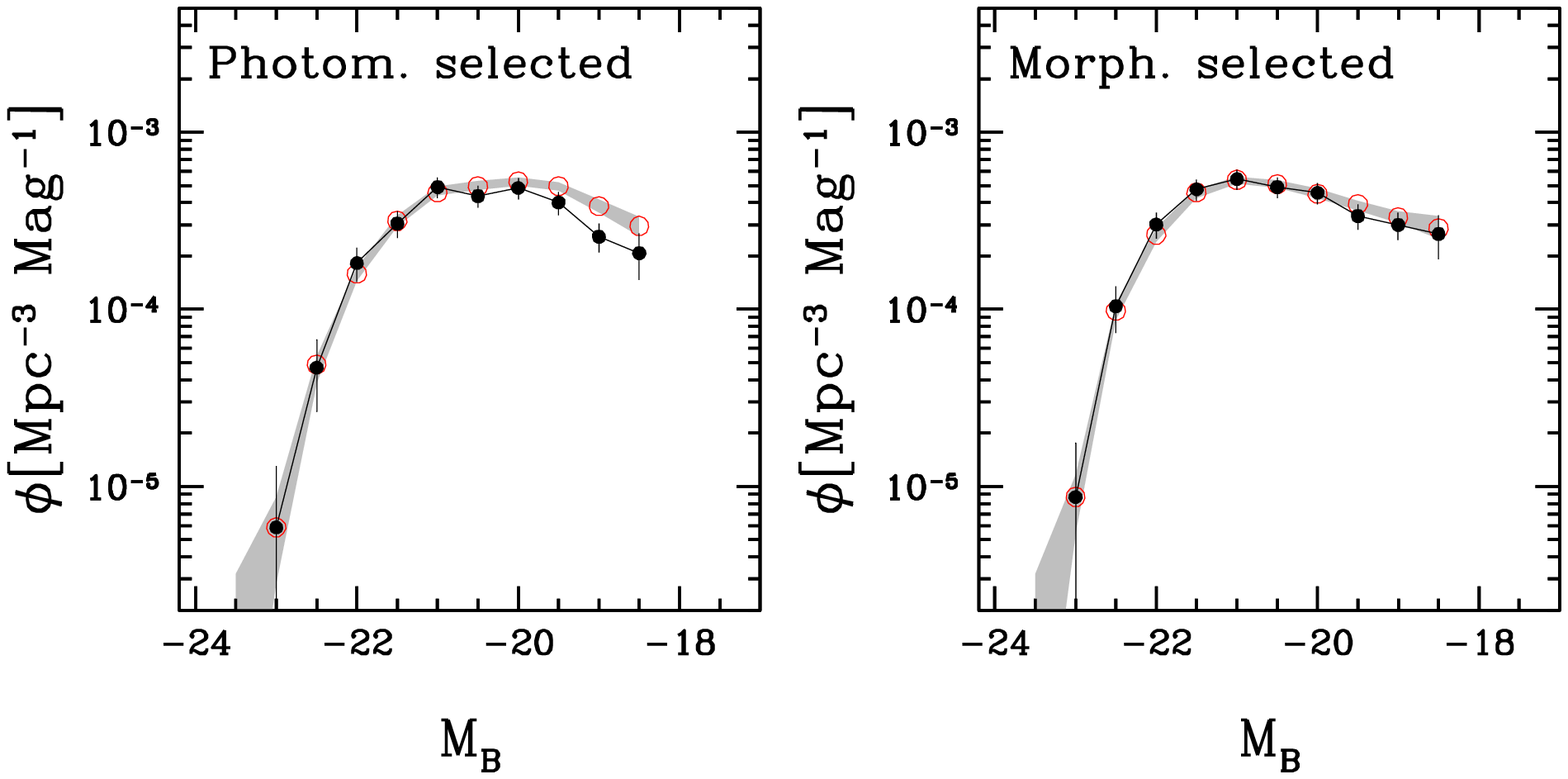}
\caption{Results of the simulations performed to assess the impact of
  the photometric redshift errors in the calculation of the LFs in the
  redshift range $0.6<z\le 0.8$.  The solid circles with errorbars
  represent the LF of the original COSMOS catalog used to generate the
  100 simulated datasets. On the right and left panel we show,
  respectively, the results for the photometrically- and
  morphologically-selected samples of early type galaxies.  The grey
  area indicates the 16$^{th}$-to $84^{th}$ percentiles of the
  distributions -in each magnitude bin- of the 100 realization of the
  COSMOS catalog.}
\label{fig:photozsim}
\end{figure*}

In Figure~\ref{fig:photozsim} we present the results of this test in
the redshift range $0.6<z\le 0.8$. This is the redshift range we use
to derive the main conclusions of our paper.  In
Figure~\ref{fig:photozsim} solid circles with errorbars represents the
LF computed using the ``true'' photometric redshifts, and the open red
circles represent the median of the 100 realizations.  The shaded grey
area associated with the simulated volume densities represent the
16$^{th}$ and 84$^{th}$ percentiles of the simulated distributions
within each magnitude bin.  Figure~\ref{fig:photozsim}  shows
that the dominant effect of the redshift uncertainty is at the faint
end of the LF, where the number of galaxies is sistematically higher
in the error-convolved LFs. This effect is stronger in the color
selected samples than in the morphologically selected one.

At the bright end, the only significant effect is for very bright
magnitudes (M$_B< -23.$, i.e., $L>10L_*$), where we would overestimate
the number of galaxies. Indeed, in the simulations we find $1-2$ galaxies with
these magnitudes in 8 out of the 100 mock realizations. 
These resultss are consistent with those of  Brown et al. (2007) who
performed similar simulations  and find that the effect of photometric
 redshfit errors on the LF is only significant for $M_B -5\log(h)<-22.5$. 
We note however, that no such galaxies are observed in the original 
COSMOS catalog. These tests show that, for the area
covered by the present analysis, Poissonian errors dominate the
contribution in the error budget at the very bright end of the LFs.

The  simulations do not include catastrofic redshift failures ( $\Delta  
z=z_{\rm phot}-z_{\rm spec}>5\sigma$), that are found to amount to 
$\sim 1$\% down to $I_{AB}=22.5$, and $\sim 2$\% down to $I_{AB}=24$. 
Although a small fraction, catastrofic failures could have a 
significant effect in the poorly populated bright part of the LF.  
However, the typical redshift
degeneracy observed for the ``training'' sample of galaxies with
spectroscopic redshift is between $z\sim 0.2$ and $z\sim 2.8$
\citep[see ][]{feldmann2006}, i.e., outside of the redshift range
considered in this study.  We also note that our conclusions are based
on galaxies with observed magnitudes brighter than $I_{AB}=21$; at
these magnitudes our photometric redshifts are very accurate.

The main conclusion of our paper is based on the number density
  of galaxies brighter than 2.5$L*$, in the redshift range $0.6<z\le
  0.8$. The main question to ask is whether the red ETGs galaxies
  brighter than 2.5$L*$ are really in the redshift range
  $0.6<z\le0.8$, or have been scattered there because of photometric
  redshift errors.  For 18 ETGs brighter than 2.5L*, spectroscopic
  redsfhit is available from the current version of the zCOSMOS-bright
  (i.e.,$I_{AB}\le 22.5$) catalog (Lilly et al. 2007): 18/18 have
  spectroscopic redshift in the 0.6--0.8 range.  This, therefore,
  demonstrate that the our result is robust against photometric
  redshift errors.

\section{Comparison with redshift  zero: 1813 SDSS galaxies}
\label{app:sdss}

We use the set of artificial images created by \citet{kampczyk2006}
from a selected sample of SDSS galaxies in order to calibrate our
results with the $z=0$ Universe.  The artificial images simulate how
the local galaxies would be observed in the COSMOS survey at redshift
$z=0.7$.  Details on the generation of these images are given in the
original reference.

In brief, the SDSS galaxies were selected in the redshift range
$z\in[0.015, 0,025]$ to be brighter than $M_B=-18.5$. The SDSS
$g$-band images of the selected galaxies were transformed to F814W
COSMOS ACS images at $z = 0.7$. At this redshift, the redshifted $g$
band well matches the HST F814W passband; thus, the redshifting of the
SDSS galaxies needs only to take care of the different pixel scales
and point spread functions, and of cosmological surface brightness
dimming.  No galaxy size evolution was considered. The $z = 0.7$
simulated SDSS galaxies were then added into the COSMOS ACS images to
reproduce the same circumstances of image crowding, noise, and so on.
We expect the SDSS sample to be representative of the local galaxy
population down to $M_{B}=-18.5$ since, overall, the
photometric$+$spectroscopic SDSS data are mostly complete down to
$r\sim 17.8$ \citep{strauss2002}, i.e., well below the considered
absolute magnitude cut that we applied.

\begin{figure*}[ht]
\epsscale{1.} 
\includegraphics[width=9cm]{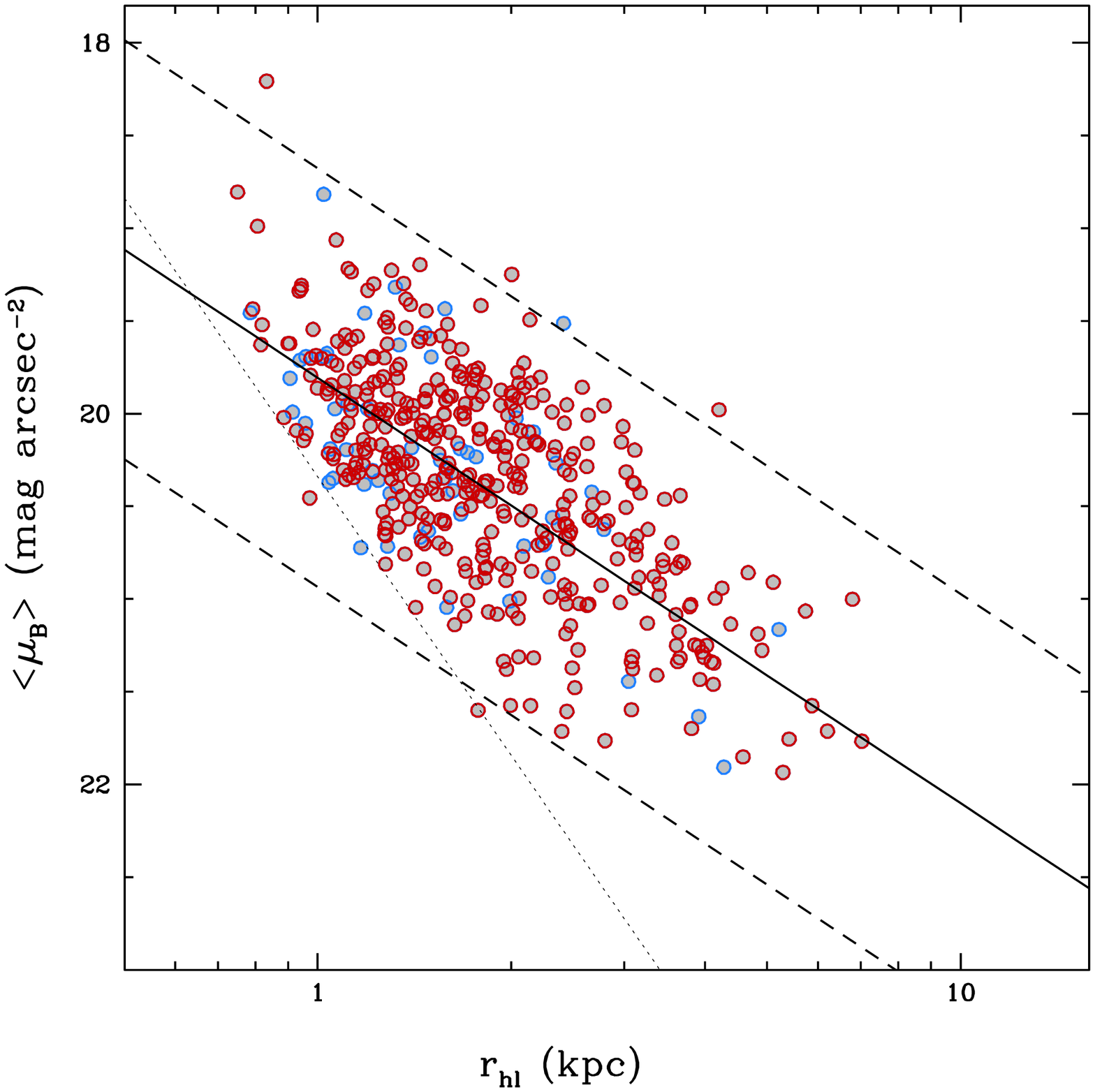}
\includegraphics[width=9cm]{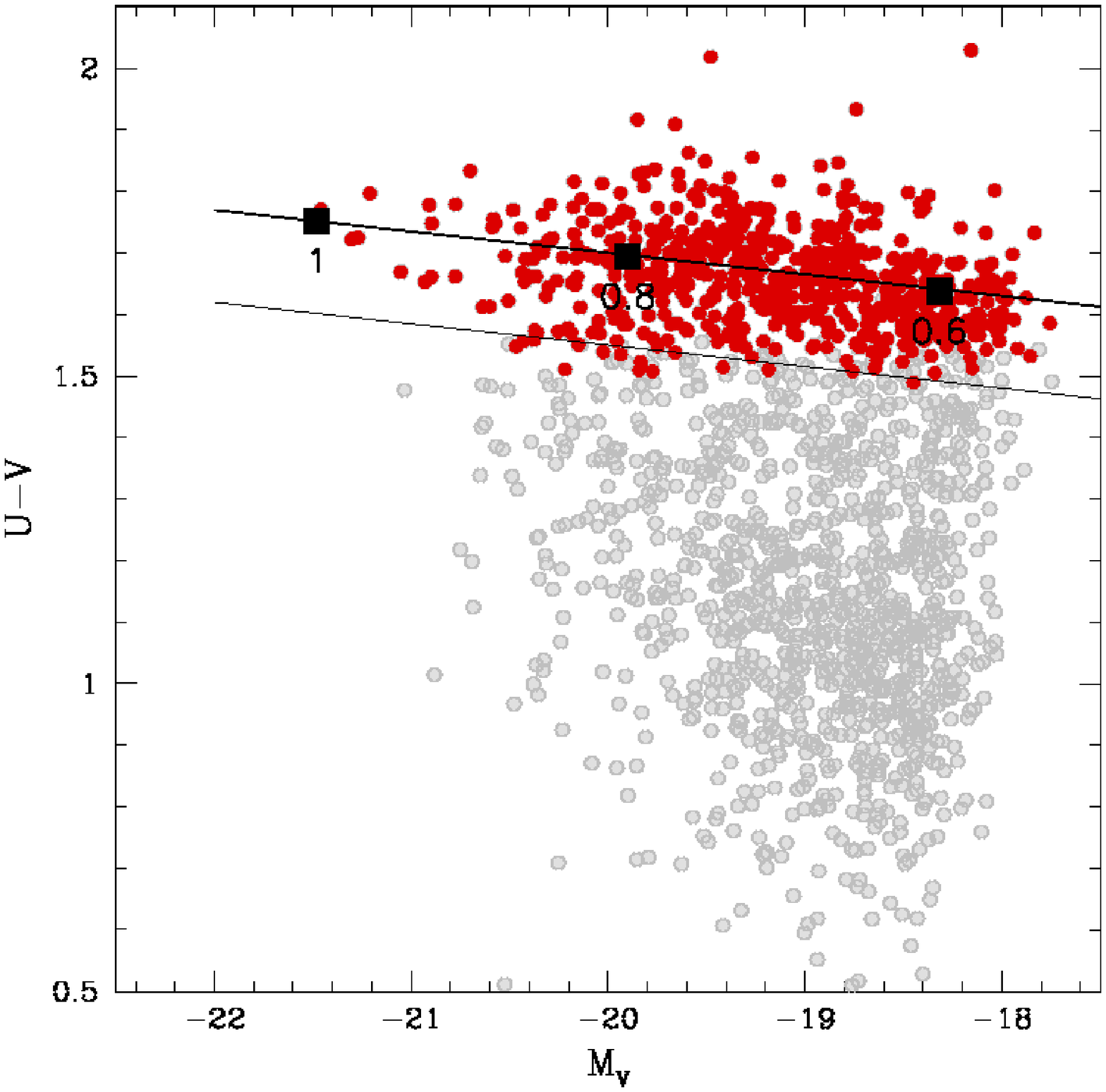}
\caption{Left panel: Kormendy diagram for all SDSS galaxies classified
  as Type~1 (early-type) and Type~2.0 (bulge-dominated disk galaxies) 
with ZEST.  Points are color--coded according
  to their ZEBRA photometric class: red points correspond to galaxies
  with ZEBRA elliptical-galaxy fits, and blue points are used for
  galaxies with SEDs fitted by ZEBRA with later spectral types. The
  solid black line shows the best linear fit derived for Coma cluster
  galaxies (J\o rgensen, Franx \& Kj\ae rgaard 1995).  Dashed lines
  are $2\times{\rm rms}$ away from the COMA best--fit line. The dotted
  line shows the magnitude limit applied to select the SDSS sample.
  Right panel: Rest-frame $U-V$ versus the absolute $V$ magnitude for
  all SDSS galaxies (gray circles). In red are highlighted galaxies
  with an SED best--fitted by ZEBRA with an elliptical-galaxy template.
  The large black squares show the aperture-corrected colors and
  magnitudes for a set of single stellar population models with
  different metallicities (from $0.6Z_{\odot}$ to $Z_{\odot}$).}
\label{fig:sdss}
\end{figure*}

The $z=0.7$ simulated SDSS galaxies were analyzed with the same
procedure used for the real COSMOS data. First, we run SExtractor to
perform the detection and measure the $I-$band magnitude, position
angle and elongation. For the detection we used the same SExtractor
configuration parameters used to generate the COSMOS--catalog
\citep{leauthaud2006}. A fraction of 6\% of the simulated galaxies was
not detected, while a fraction of 0.1\% was found with $I\gtrsim 24$,
and therefore excluded from the sample.  We then classified the
galaxies using ZEST. Rest-frame properties such as $U-V$ colors,
$B-$band absolute magnitudes, radii in kpc, were derived assuming that
all galaxies were at redshift $z=0.7$; the $U-V$ colors and $M_V$
magnitudes were derived by logarithmic interpolation between the
observed SDSS magnitudes ($u$, $g$, and $r$).

In Figure~\ref{fig:sdss} we present the $U-V$-$M_V$ color--magnitude
diagram (right panel) and the Kormendy diagram (left panel) for the
simulated $z=0.7$ SDSS galaxies. In particular, we show on the
Kormendy diagram all galaxies classified as Type~1 or Type~2.0 with
different colors according to the ZEBRA best--fits to their SEDs.
As in Figure~\ref{fig:kormendy}, red points represent
elliptical-galaxy SEDs, and blue points represent later photometric
types. As expected, the SDSS morphologically-classified ellipticals
are all consistent with the $z=0$ Kormendy relation, as possible
small-size faint interlopers are excluded from the sample by the
imposed magnitude cut (see dotted line in the left panel of
Figure~\ref{fig:sdss}, which shows the $I_{AB}=24$ magnitude limit).

In the {\it right panel} of Figure~\ref{fig:sdss} we show as gray
points all SDSS galaxies in the original sample, and, highlighted in
red, all galaxies with a ZEBRA elliptical-galaxy SED. The rest--frame
colors were derived using the SDSS total magnitudes, and therefore
represent the global galaxy color. As mentioned in
Section~\ref{sec:slope}, ETGs show color gradients
with an average value of ${\rm d}(U-V)/{\rm d}(\log{R})=-0.15$
\citep{scodeggio2001}. Due to this color gradient, the galaxy colors
depend on the aperture size used for the photometric measurements; the
colors are systematically bluer the larger the aperture. Since the
average color gradient does not correlate with galaxy luminosity, the
global effect of a large aperture is a blueing of the red--sequence
relative to that derived using a smaller aperture. Assuming, for
simplicity, that early--type galaxies in the local universe have a De
Vaucouleur profile, and using the color gradients published by
\citet{scodeggio2001}, we estimate that the difference between the
$U-V$ colors computed within 30\% of the half-light radius and those
calculated using total magnitudes amounts to $\sim 0.1$ magnitudes.
In the Figure, the colors are corrected to account for the
aperture corrections.

Assuming that the color--magnitude relation of $z=0$ early--type
galaxies can be interpreted as a pure metallicity sequence
\citep{kodama1998,bernardi2003d}, with less luminous galaxies being
more metal poor than more luminous galaxies, in Figure~\ref{fig:sdss}
we show the $U-V$ colors derived for single stellar population models
of different metallicities formed at $z_f=2$ \citep{bc03}. The colors
of the models, together with the absolute $V-$band absolute magnitude,
metallicity and mass of the galaxy are reported in Table~\ref{tbl:1}.

\begin{center}
\begin{deluxetable}{cccc}
\tabletypesize{\scriptsize}
\tablecaption{Models of elliptical galaxies at $z=0$ ($z_f=2$).\label{tbl:1}}
\tablewidth{0pt}
\tablehead{
\colhead{$M_V$} & 
\colhead{$U-V$} & 
\colhead{$Z/Z_{\odot}$} & 
\colhead{$M_{*}$} \\
\colhead{(mag)} & 
\colhead{(mag)} & 
\colhead{} & 
\colhead{($M_{\odot}$)} \\
\colhead{(1)} & 
\colhead{(2)} & 
\colhead{(3)} & 
\colhead{(4)} }
\startdata
 $-$18.3 &  1.75   &   0.62  &  9.5$\times 10^{9}$ \\
 $-$19.9 &  1.81   &   0.79  &  4.4$\times 10^{10}$ \\
 $-$21.5 &  1.87   &   0.99  &  2.0$\times 10^{11}$ \\
 $-$23.4 &  1.93   &   1.26  &  1.2$\times 10^{12}$ \\
\enddata
\tablecomments{The columns are: (1) $V-$band absolute
  magnitude; (2) $U-V$ rest-frame color, uncorrected for aperture
  effects; (3) stellar metallicity;  and (4) total stellar mass.}
\end{deluxetable}
\end{center}

\acknowledgments{We thank the anonymous referee for the careful
  reading of the manuscript, and the comments that helped to improve
  the presentation and discussion of our results. The HST COSMOS
  Treasury program was supported through NASA grant HST--GO--09822.
  We gratefully acknowledge the contributions of the entire COSMOS
  collaboration.  More information
  on the COSMOS survey is available at 
  http://www.astro.caltech.edu/$\sim$cosmos. \\We thank the NASA
  IPAC/IRSA staff for providing online archive and server capabilities
  for the COSMOS datasets.  T.  Lisker is acknowledged for helping
  with a preliminary reduction of a fraction of the ground-based
  near-infrared data. CS acknowledges support from the Swiss National
  Science Fondation.}

{\it Facilities:} \facility{HST (ACS)}

 \end{document}